\documentclass[rmp,aps,nofootinbib,twocolumn,floatfix,superscriptaddress]{revtex4}
\usepackage{natbib}
\usepackage{graphicx}
\usepackage{subfigure}
\usepackage{amssymb,amsmath}
\usepackage{mathrsfs}
\usepackage{epstopdf}
\usepackage[usenames]{color}

\def\RP{\mathbb{RP}}
\def\Z{\mathbb{Z}}

\def\T{\mathbb{T}}

\begin{document}

\title{Disclination Loops, Hedgehogs, and All That}
\author{Gareth P. Alexander}
\affiliation{Department of Physics \& Astronomy, University of Pennsylvania, 209 South 33rd Street, Philadelphia PA 19104, U.S.A.}
\affiliation{Centre for Complexity Science, Zeeman Building, University of Warwick, Coventry, CV4 7AL, U.K.}
\author{Bryan Gin-ge Chen}
\affiliation{Department of Physics \& Astronomy, University of Pennsylvania, 209 South 33rd Street, Philadelphia PA 19104, U.S.A.}
\author{Elisabetta A. Matsumoto}
\affiliation{Department of Physics \& Astronomy, University of Pennsylvania, 209 South 33rd Street, Philadelphia PA 19104, U.S.A.}
\affiliation{Princeton Center for Theoretical Science, Princeton University, Princeton, NJ 08544, U.S.A.}
\author{Randall D. Kamien}
\email{kamien@physics.upenn.edu}
\affiliation{Department of Physics \& Astronomy, University of Pennsylvania, 209 South 33rd Street, Philadelphia PA 19104, U.S.A.}

\date{\today}

\begin{abstract}
The homotopy theory of topological defects is a powerful tool for organizing and unifying many ideas across a broad range of physical systems.  Recently, experimental progress has been made in controlling and measuring colloidal inclusions in liquid crystalline phases.  The topological structure of these systems is quite rich but, at the same time, subtle.  Motivated by experiment and the power of topological reasoning, we review and expound upon the classification of defects in uniaxial nematic liquid crystals.  Particular attention is paid to the ambiguities that arise in these systems, which have no counterpart in the much-storied XY model or the Heisenberg ferromagnet. 
\end{abstract}
\maketitle
\tableofcontents

\section{Introduction}
\label{sec:introduction}

First identified through the beautiful textures of their defects, liquid crystalline materials may very well be the ideal proving grounds for exploring notions of broken symmetries, associated Goldstone modes, phenomena akin to the Higgs mechanism, and, of course, topological defects.  Indeed, the subjects of topological defects and liquid crystals are so intertwined that it is difficult to see how a thorough understanding of one could be garnered without knowledge of the other.  Moreover, it has proven fruitful to treat the defects as fundamental excitations;  solitons in the sine-Gordon model can be treated as interacting fermions~\cite{coleman75}, vortices in the XY model drive the Kosterlitz-Thouless phase transition~\cite{kosterlitz73}, controlling flux lines is essential for maintaining superconductivity \cite{nelson88}, and the theory of magnetic monopoles allows them to be viewed as point particles \cite{colemanbook}.  In all of these cases, the fluctuations and configurations of the smooth background field around the defects can be replaced with an effective interaction between the defects.  Thus the elasticity and statistical mechanics of these systems can be recast in terms of a discrete set of topological charges. Because of these successes and physicist's deeply-ingrained love for Gauss's law and the connection between flux and charge, we naturally attempt to interpret more complex systems in the same terms.  Unfortunately, these two cases are exceptions -- in this colloquium we will emphasize this important point and attempt to clarify if, when, and how it is appropriate to view defects as point charges.

Nematics are a prototypical liquid crystalline material -- though a
sample will flow like a liquid, owing to the lack of positional order,
the rod-shaped, typically organic or aromatic ring molecules are locally aligned, leading to
orientational order, with widely-applied optical consequences. 
Defects in a nematic liquid crystal, as in any ordered medium, are places where the order changes discontinuously and is thus ill-defined. They occur in the form of isolated points and lines, either system spanning with endpoints on the boundaries of the sample or closed up into loops. We will use as our example the three-dimensional uniaxial nematic, a system that turns out not to be as simple as it might seem.  Not only is it, in some sense, the simplest counterexample to the ``defects as charges'' approach, but it is also now extremely relevant from the experimental point of view as it has been demonstrated that defects can be manipulated in nematic cells with great precision and variety allowing complex links and knots to be tied \cite{tkalec11,kamien11}.
Though the
history or path dependence of the defect motion at once spoils
the program of reducing the configuration space to discrete data, it
simultaneously introduces a new set of topological degrees of
freedom in the three-dimensional nematic.  The path-dependence imposes a non-Abelian structure to the classification of states.   In other liquid
crystalline systems, line defects in the biaxial nematic are
anticipated to entangle topologically~\cite{poenaru77} and the
interaction between disclinations and dislocations in smectics suffers
similar path dependence~\cite{poenaru77,chen09}.  Certain non-Abelian quantum systems have been proposed as a route to
topological codes and computation~\cite{kitaev03,nayak08}.

Although they arise naturally during phase transitions, defects can also be induced deliberately by local excitation with a laser~\cite{smalyukh09} or through boundary conditions as with colloidal inclusions where the anchoring of molecules at the colloid surface induces the presence of defects in the surrounding liquid crystal. This is an especially rich technique allowing for the generation of a wide range of defects and for studying the interactions between them. For instance, the defect accompanying a colloid may take the form of an isolated point~\cite{poulin97} or of a disclination loop encircling the particle in a `Saturn ring' configuration~\cite{terentjev95}, while two or more colloids in close proximity can form a variety of `entangled' structures where a single disclination loop wraps around both of them~\cite{guzman03,ravnik07,ravnik09}. These last situations realize the interaction between line and point defects that is a central aspect of this colloquium. 

In Section \ref{sec:lines} we begin our tale by reviewing the
classification of topological defects by looking at point defects in two-dimensional
nematics, and line defects in three-dimensional nematics.  In Section
\ref{sec:hedgehogs} we continue to point defects in three-dimensional
nematics.  With all the characters introduced, we proceed to
discuss our main point: that there are only two types of unlinked disclination loops.
What we present is built on the classic review by Mermin on topological defects~\cite{mermin79} and a definitive result by J\"anich~\cite{janich87} which finds the topological classes of uniaxial nematic textures on $\mathbb{S}^3$, the three-dimensional sphere, (equivalently, three dimensional space with fixed boundary conditions at infinity), with point and line defects.  We will also discuss the biaxial nematic phase and show that it affords additional physical insight into the analysis of uniaxial defects.  

\section{Circular loops around defects: the fundamental group}
\label{sec:lines}

\begin{figure}[!h]
\begin{center}
\includegraphics[width=3.25in]{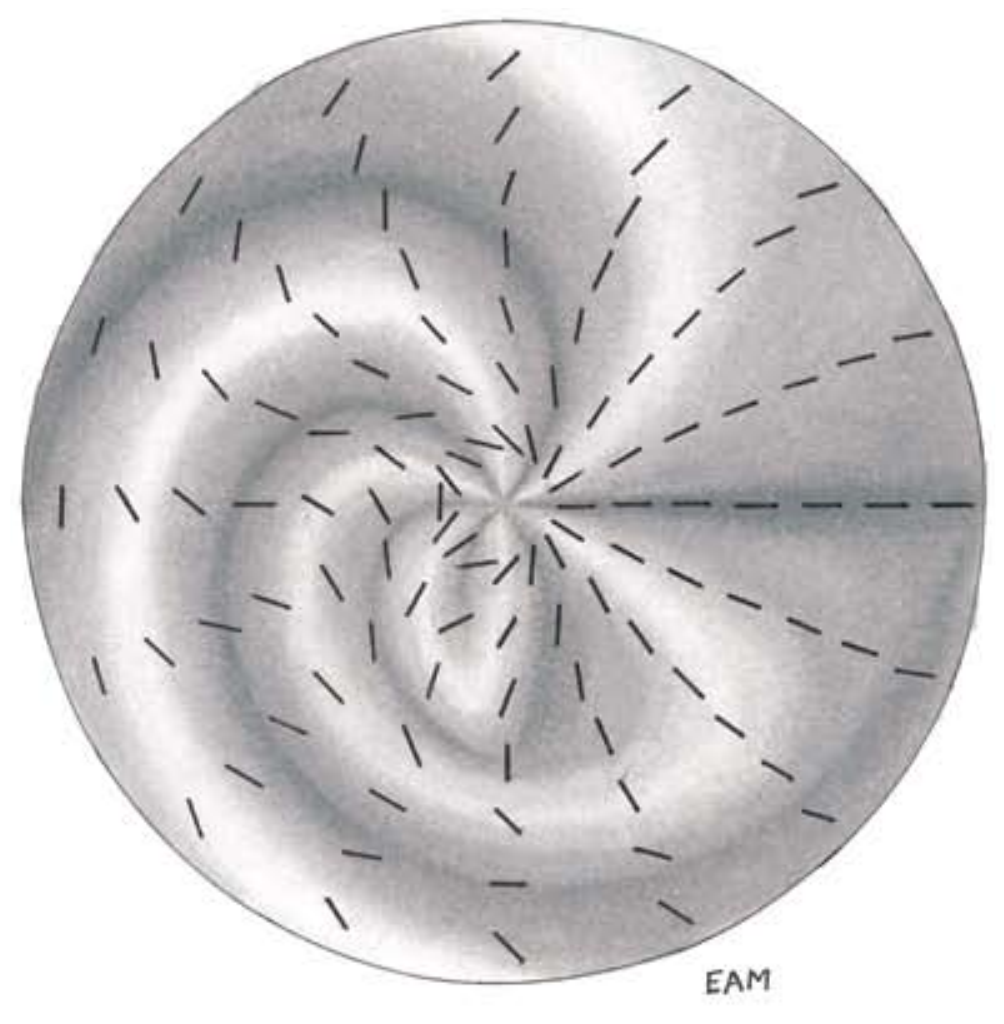}
\caption{Cartoon of the director field and the resulting brush pattern corresponding to the homotopy in equation (4).  The radial coordinate $t$ parametrizes the sequence of maps $f_t(\phi)$.  Near the defect center, six brushes are visible so that simply counting would give a winding of $\frac{3}{2}$. However, away from the defect core, the texture smoothly deforms to a texture that may be seen to have a winding of $\frac{1}{2}$.}
\label{fig:brushes}
\end{center}
\end{figure}

A common and distinctive experimental procedure for imaging the orientational order is to place the sample between crossed polarizers, yielding in nematic materials the characteristic Schlieren texture from which they derived their name.  In FIG. \ref{fig:brushes} our eyes are drawn to the dark brushes and tend to follow them towards
their intersections.  The brushes typically meet at points in twos or 
fours; larger numbers are possible but they too are multiples
of two. The dark brushes correspond to regions where the orientation is parallel to either of the mutually perpendicular polarizer or analyzer directions
and the points at which they meet are disclination defects. 
Why defect?
Because there the orientation is ill-defined, as the brushes would tell us to assign two
or more different orientations to that point.  
We can fix this by removing the defect point by poking a hole in the material.
Physically, of course, we don't have a hole in our sample, just a
place where we don't know the order.  The problem is fixed by letting
the magnitude of the order vanish at the origin.  Then no hole is needed, but there is a point where
there is no
longer nematic order and the core is in the isotropic phase. To the
reader acquainted with defects in superfluids and superconductors,
this should sound familiar: in the center of an Abrikosov vortex in a
superconductor \cite{Abrikosov} the superconducting order parameter
vanishes and there is normal metal -- a hole in the superconductor
\cite{CL}.  In all these cases we can detect the
existence of a defect without any detailed knowledge of the disordered
core: Topological defects are characterized by the boundary
conditions at the interface between the higher symmetry state and the
lower symmetry state.  In the Schlieren texture, these boundary
conditions may be read off from the pattern of brushes.

In this section we introduce the basic concepts in the study of
topological defects in this simplest case and connect the observation of brushes to the presence of defects.  We will cover what defects
are, the use of measuring surfaces to probe them, and finally how to
combine them.
We begin by recalling that the continuum description of uniaxial nematics is based upon the assignment of a local `average molecular orientation' to every point in the sample, called the director field.
 This average orientation is that of a rod-like object, rather than an arrow, so that the nematic director is properly a line field, rather than a vector field. Moreover, it is only the orientation of the director that is important, not its magnitude, so that we may take the director to be a unit vector ${\bf n}$, subject to the condition that ${\bf n}$ and $-{\bf n}$ are identified. 

\subsection{Two-dimensional Director: $\mathbb{RP}^1$}
\label{subsec:2d}

First we consider nematic samples in which the director always lies in the $xy$-plane, a prototypical thin cell situation. 
Equivalently, we consider the projection of a three-dimensional director onto the $xy$-plane:
\begin{equation}
{\bf p} = A(x,y)\bigl[ \cos\left(\theta(x,y)\right),\sin\left(\theta(x,y)\right)\bigr],
\end{equation}
where $A(x,y)$ is the nonvanishing magnitude of the projection and $\theta(x,y)$ is
the angle it makes with the $\hat x$-axis.  Just as ${\bf n}$ and $-{\bf
n}$ are identified, so too are $\bf p$ and $-\bf p$ or, equivalently,
$\theta(x,y)$ and $\theta(x,y)+\pi$.  
Because the polarizer and analyzer directions are $\pi/2$ apart, each dark brush around a defect point tells us about a $\pi/2$ rotation of the director. The number of brushes counts the winding of the director field.  The fact that we have defects with two brushes directly tells us that rotations by $\pi$ leave the phase invariant and is experimental proof of the symmetry of the nematic phase~\cite{frank58}. The angle that the
director turns on a loop around a defect divided by the angle $2\pi$ of a
full turn, is commonly referred to as the strength, or charge, of the disclination and is equal to $\pm 1/4$ times 
the number of brushes.  
Though the absolute value of the strength may easily be extracted from counting the number
of brushes, the sign of the winding is not yet determined.
It is a pleasant exercise in visualization to show that upon rotating
the polarizers the brushes will counter-rotate when the defect has
negative winding and will co-rotate when the winding is positive.

As defects are points of {\sl discontinuity} of the director field,
we aim for a classification up to {\sl continuous deformations} of their
surroundings. 
When two paths can be continuously distorted
into each other, they are said to be (freely) {\sl homotopic}. It is useful to
think of these deformations as time evolution because it not only
allows us to visualize the distortion but it also reminds us that
under continuous dynamics, configurations remain in the same 
{\sl homotopy class}. 
Above we began to think about a defect in terms of a measuring loop
around it -- this key idea is what allows us to probe the properties of a singularity in our
texture without leaving the safety of the well-behaved director field.
Let's consider the possibilities for the director on a small round
measuring circle in the system.  Here's an explicit family of possible
measurements on this circle ${\bf p}/A={\bf c}_m$ which exhibit any even 
number of brushes:
\begin{equation}
{\bf c}_m = \bigl[ \cos(m\phi), \, \sin(m\phi) \bigr] ,
\label{eq:schlieren}
\end{equation}
with $m \in \tfrac{1}{2} \mathbb{Z}$ a half-integer and $\phi$ the polar
angle around our circle.
These textures are all distinct in that we cannot continuously deform 
${\bf c}_n$ into ${\bf c}_m$ for $n\ne m$.  Thus within this family, the half-integer
$m$ classifies the possibilities for the director field.
Of course the textures described by \eqref{eq:schlieren} are special
choices, or simplifications, and
the director is rarely of this exact form even very close to the
defect line.  
However, any director field that we might
have measured on our circle can be homotoped into one of
these! 
The utility of \eqref{eq:schlieren}, therefore, is that these fields 
capture the possible windings of a general director field, and thereby
form a
set of representatives for the behavior outside point defects.  Most placements of our
measuring circle will not surround a defect and hence the director field 
can be deformed to ${\bf c}_0$, which is a constant.  On the other
hand, if the measuring circle surrounds a point where two brushes
meet, it should be homotopic to ${\bf c}_{1/2}$ or ${\bf
c}_{-1/2}$. 

Though reasonable sounding, the previously stated connection between
brush counting and winding number relies on a hidden assumption!
Suppose ${\bf p}=[\cos f(\phi),\sin f(\phi)]$ on our measuring circle
and $f$ took the form
\begin{equation}
f(\phi) = \begin{cases}\frac{3}{2}\phi&\quad 0\le\phi<\frac{2\pi}{3},\\  2\pi -\frac{3}{2}\phi&\quad \frac{2\pi}{3}\le\phi<\frac{4\pi}{3},\\ \frac{3}{2}\phi - 2\pi&\quad \frac{4\pi}{3}\le\phi<{2\pi},\\ \end{cases}
\end{equation}
where we first go from $0$ to $\pi$, back to $0$ and then back to
$\pi$.  Now $f$ takes values in the interval $[0,\pi]$ with
0 and $\pi$ identified due
to the ${\bf p}\rightarrow{-\bf p}$ symmetry. Were we to count the number of times we landed at $\pi/2$ for
instance, we would find $3$, as we would for any generic point of
$[0,\pi]$.  Under crossed polarizers we would see six brushes.  But we can also smoothly distort the map $f(\phi)$ so that there are only two brushes through the sequence of maps $f_t(\phi)$, $t\in[0,1]$:
\begin{equation}
f_t(\phi) = \begin{cases}\left(\frac{3}{2}-t\right)\phi& 0\le\phi<\frac{2\pi}{3-2t},\\  2\pi -(\frac{3}{2}-t)\phi& \frac{2\pi}{3-2t}\le\phi<\frac{(4-2t)\pi}{3-2t},\\ (\frac{3}{2}-t)\phi +(2t- 2)\pi& \frac{(4-2t)\pi}{3-2t}\le\phi<2\pi.\\ \end{cases}
\end{equation}
We see that $f_0(\phi)=f(\phi)$ while $f_1(\phi)$ generates the charge
$1/2$ map ${\bf c}_{1/2}$ as $\phi$ ranges from $0$ to $2\pi$.  So
have we shown that charge $3/2$ is the same as charge $1/2$?
Certainly not!  We have learned, rather, that we must be more careful
about counting brushes.  We assumed in our previous rule of counting brushes to determine angles that the angle function was always either increasing or
decreasing on our loop.
In general, when counting the number of times we pass $\pi/2$ we must
keep track of the way we pass it -- is the angle increasing or
decreasing?  
We thus look at the net rotation
of $f(\phi)$ by calculating the {\sl winding number} via the following
integral
\begin{equation}
w = \frac{1}{2\pi}\int_0^{2\pi}\! d\phi \frac{df_t}{d\phi} = \frac{1}{2} ,
\label{eq:winding}
\end{equation}
for all $t$.  In general, the winding number is precisely the charge
of the defect.  In actual liquid crystal systems, energetics strongly
disfavors extra brushes because these would correspond to regions of
large splay or bend in a nematic.  Thus, the na\"ive counting of
brushes works outside of exceptional circumstances.
Though we have framed the discussion above in terms of a particular 
measuring circle, the classification by winding numbers holds for
any measuring loop disjoint from the defects --
just let $\phi$ be any coordinate which goes from 0 to $2\pi$ as we
traverse the loop once.  In particular, the texture on any loop in the 
sample which lassos just one of the defects is homotopic
to any other; this can also be seen by the fact that the number of
brushes meeting at the defect is conserved.
Similarly, the number of brushes meeting at a point
in a Schlieren texture remains unchanged in time as 
the texture evolves except when two defects combine and annihilate
each other. 
The winding number $m$ is a key
example of an essential idea in topology: {\sl turning geometric
information into counting}.  By doing so we get robust measures since,
in particular, integers (or half-integers), being discretely valued,
cannot change continuously.  

\subsection{Making Precise Measurements}
\label{subsec:precise_measurements}

Let's start putting our above classification into a broader context.  We have
mentioned several different topological spaces which we should keep
straight.
First, we are dealing with uniaxial nematic textures, so there is a
space associated with the possible ground states of the director
field, the {\sl ground state manifold} (GSM)\footnote{Recall that the
ground states of a nematic are just uniform textures with the director
pointing in some direction, so that in a general texture the local
orientation can always be identified with one of these ground states.
Thus the changing orientation of the director in a general texture may
be thought of as motion on the ground state manifold.  There is not an
accepted term for this space in the literature, where it has also been
called the {\sl manifold of internal states} and the {\sl order
parameter space}.}. 
Here, the angle of the director $\Upsilon$ lives in the interval $[0,\pi]$ with $0$
and $\pi$ identified.  This space is known as the real projective
line, $\mathbb{RP}^1$. Like any interval of the real line with
endpoints identified, it has the topology of a circle.
Next, we have the space $M$ making up the sample volume.  For our
Schlieren textures, $M$ may be taken to be a suitable region of the plane.  Inside $M$ there
is a 
subspace, the set of defects $\Sigma$ where the order is not 
defined.  In the last section, $\Sigma$ was the set of disclination points.
We obviously have the director field, an $\mathbb{RP}^1$-valued field
on $M$ away from the defect set, ${\bf n}: \, M \setminus \Sigma
\rightarrow\text{GSM}$. In general, it is daunting to contemplate the texture on the entire sample 
especially if there are many defects.  We also would like to understand to what
extent ${\bf n}$ may be understood as arising from its behavior near
the defect set -- might there be a way to cut $M\setminus\Sigma$ into
more manageable pieces around its defects, classify those, and then
glue them back together?
These considerations motivate the introduction of auxiliary spaces: measuring surfaces of fixed 
topology contained in the sample.  We study the behavior of the director field
on these as we vary their placement relative to the defect set or as
we vary ${\bf n}$.  
Indeed, all the topological ``charges'' in this paper are defined with respect to a choice of measuring surface. Saying that a defect carries a charge is really shorthand for saying that a small measuring surface which surrounds only that defect measures such a charge. If such a measuring surface is not  present, then the ``charge of a defect'' is ill-defined.
Of course, these ambiguities are nothing new and have been discussed before \cite{mermin79}.  However, with the interest in nematics and cholesterics with embedded colloids \cite{musevic06,lintuvuori10,copar11}, these mathematical subtleties are no longer just about precision or axiomatic rigor -- they are absolutely necessary for the proper interpretation of data.  
Above we measured the director on loops (which are topologically
circles, denoted by $\mathbb{S}^1$)  and
observed that the winding number $m$ was constant during deformations
of the loop or the texture as long as no defects pass through the
loop, which essentially solved our classification problem for single defects
in Schlieren textures.  In more mathematical jargon, we saw that
the set of maps from $\mathbb{S}^1$ to
$\mathbb{RP}^1$ up to homotopy (denoted
$[\mathbb{S}^1,\mathbb{RP}^1]$) was equivalent to the set of
half-integers $\frac{1}{2}\mathbb{Z}$.

How can we go from single defects back to the full texture?
Let's draw a picture of a two dimensional plane with punctures at the
defects and small measuring loops about each puncture.  Given the
winding numbers at the loops, do we know the classification of the
full texture?  This picture is
actually no different from one we might draw for the following
situation: Consider a hypothetical Schlieren texture that contains several nearby strength $\pm 1/2$
defects. If we take a measurement around a circuit
that surrounds all the defects, what do we get?  Before we say anything 
more about the local-to-global question, let us try to find an answer 
to this natural question: how
can the theory capture the intuitive notion that defects can combine
or split?  
We almost have it: we know that measuring circuits
around defects are in correspondence with half-integers, and it happens
to be true that any two half-integers can be added or subtracted.
There is just a tiny gap between question and answer now.
For one, we need to define a way to go from two measuring circuits to
one.  The natural way to add loops is to concatenate them by forming a
longer loop by running over the first and
then running over the other.  But in order to do this,  the loops have to start and end at
the same point in sample or they cannot be connected.  By the same token, the two
textures traced out on the loops must start and end at the same point in
$\mathbb{RP}^1$!  Given two arbitrary measuring loops, 
there's no reason for the textures on them to agree at some point.
On the other hand, our figure shows that we {\sl can} draw a big
circle surrounding any pair.  Is there a way to go from two circles to
one without having to make arbitrary choices in the region between the
circles?  Yes!  Since we are interested only in properties that are preserved under continuous deformations,
we can deform the texture around one defect so that the value in $\mathbb{RP}^1$ agrees at the common base point in the sample.  
To visualize this, consider a small disc around the defect.  The homotopy proceeds radially outward by deforming the original texture at the center of the disc to the new texture at its outer edge.

This solution is inspired by the following geometrical
fact: the space between a set of non-intersecting circles in the plane 
can be continuously deformed to a set of circles joined at a single
point, called a {\sl wedge sum}, or {\sl bouquet} of circles.  
Once we have the texture on the bouquet, the concatenation of the
loops is well-defined, and we thus have a definition of
addition in our system.  In the case here, 
we may form addition by measuring around the loop which hugs the ``outside'' of the bouquet.
By restricting to sets of
loops which pass through the same {\sl base point}, our {\sl set} of
equivalence classes of maps from $\mathbb{S}^1$ to the GSM is enriched 
with the additional algebraic structure of a {\sl group}.  Note that
based homotopy is absolutely necessary here -- 
 continuous distortions of paths that preserve the base point.

A quick aside on group theory is in order.  Recall that a group requires a way of
adding its elements: we just saw that this is the concatenation of
loops.
In terms of brush counting, or 
the winding number, we need not be concerned with the details
of the rate at which we traverse the GSM, or whether we pass a brush
in the first or last half of the trip.  We need only concern ourselves
with the order in which we put the two paths together.  
This addition
of classes of paths defines the group addition which may not be
commutative but happens to be in this particular case. A group requires an 
identity: in a uniform nematic
texture the map from a loop in the sample to the GSM is always
constant, so this provides the identity under the group addition. Finally, each map
has an inverse: because we can go around the GSM in the
opposite direction by backtracking our precise path, we also have an
inverse.  
Together these properties ensure that we have a group, known as the
fundamental group, $\pi_1(\text{GSM})$ -- the set of based maps from a measuring circuit $\mathbb{S}^1$ in the sample to the GSM which are equivalent up to based homotopy.
And as the reader probably has already guessed, this
group for the case where $\text{GSM}=\mathbb{RP}^1$ is precisely the
(half-)integers. A loop that goes around only one defect in our hypothetical texture will intersect 
two brushes, giving a winding of $\pm 1/2$. A larger loop encircling both 
will yield a count of four if the defects have the same sign, or zero if 
they have opposite signs.  See Mermin's review
\cite{mermin79} for a more detailed and precise discussion of these
properties.   

Though the choice of the base point is arbitrary, its constancy is
essential and leads to many of the interesting phenomena we will
discuss. The need for a base point is precisely why standard time
zones were developed for trains.  It is just a matter of having a
single clock at one location ({\sl i.e.} one measuring circuit) to
determine the elapsed time at that spot.  Two clocks can be used to
measure elapsed time as long as they are {\sl synchronized} -- that is
why we must have a base point. 
How are general maps from $\mathbb{S}^1$ to the GSM related to $\pi_1(\text{GSM})$?
In the case of Schlieren textures,
we can relax: $[\mathbb{S}^1,\mathbb{RP}^1]$ is simply the underlying set of
$\pi_1(\mathbb{RP}^1)$.  Even here, however, we must remember that if we
want a topological charge for loops around defects that satisfies addition, we
better pin our loops on a base point!  

Let's conclude this section by sketching how what we have done tells us
how to go from local measurements around defects to the full texture.
The problem of classifying the defects on the full texture if we fix a base point is equivalent to that of classifying the texture on any  bouquet of circles to which it may be retracted\footnote{It can be proved that ``deformation retractions'' of the sample onto a smaller space don't change the properties which can be probed with the tools of homotopy \cite{hatcher02}.}. 
For each circle 
in the bouquet, we can choose any element of
$\pi_1(\mathbb{RP}^1)$, so the classification for a bouquet of $k$
circles is by a $k$-tuple of half-integers.  In this case it turns out
that this is also the answer if we allow the texture at the base point to
vary.  The complication, as we will see in the following, is that there can be
more than one way to tie together the measuring circuits.  When
$\pi_1(\text{GSM})$ is non-Abelian
this freedom of choice leads to an ambiguity in measuring the ``charge'' of a defect.  This is a major point of this review and also emphasizes a possibly obvious point: the Abelian or non-Abelian nature of defects only comes into play when there is more than one defect.  Making measurements correctly is, as always, the key to understanding the physical system.

\subsection{Three-dimensional Director: $\mathbb{RP}^2$}
\label{subsec:3dlines}

In our two-dimensional example, we insisted that the director never point perpendicular to the plane and it was then possible to describe the texture in terms of idealized configurations where the director was entirely planar. This situation is often an accurate description of thin samples where the bounding surfaces are treated to promote planar alignment.  
However, in bulk samples, the nematic director can point along any of the directions in three dimensions with important implications for the topology.  

The defects described in this subsection still may be captured by a
measuring circle in the sample -- in a three-dimensional nematic,
these are line defects.  By considering a two-dimensional
cross-section of a three-dimensional nematic, we can discuss measuring
circles around line defects 
as circles surrounding points in the plane, as in the previous
subsection.  Thus the sum of the measuring circles about two line defects 
corresponds to the process of merging two parallel defect segments into 
one.  It is not hard to see that the constructions explained in
the last subsection have direct analogues for line defects probed this way,
and we leave elaboration on most of them to the reader.  
In later sections, we will probe line defects with other
measuring surfaces.  
 
The GSM here is a sphere with antipodal points identified (as ${\bf
n}$ and $-{\bf n}$ are identified in a line field).  In this instance, the geometry is simple and we are able to see in our mind's eye the topology of the GSM.  In more general situations, it is useful to have something more systematic: indeed, without a systematic approach we can never be sure that our intuition is not fooling us. The general framework of the Landau theory of phase transitions that exploits
symmetries and symmetry breaking not only provides well-known, systematic tools to our problem, but also emphasizes the essential aspect of topological defects -- a higher symmetry phase surrounded by a lower symmetry phase.  

\begin{figure}[t]
\begin{center}
\includegraphics[width=3in]{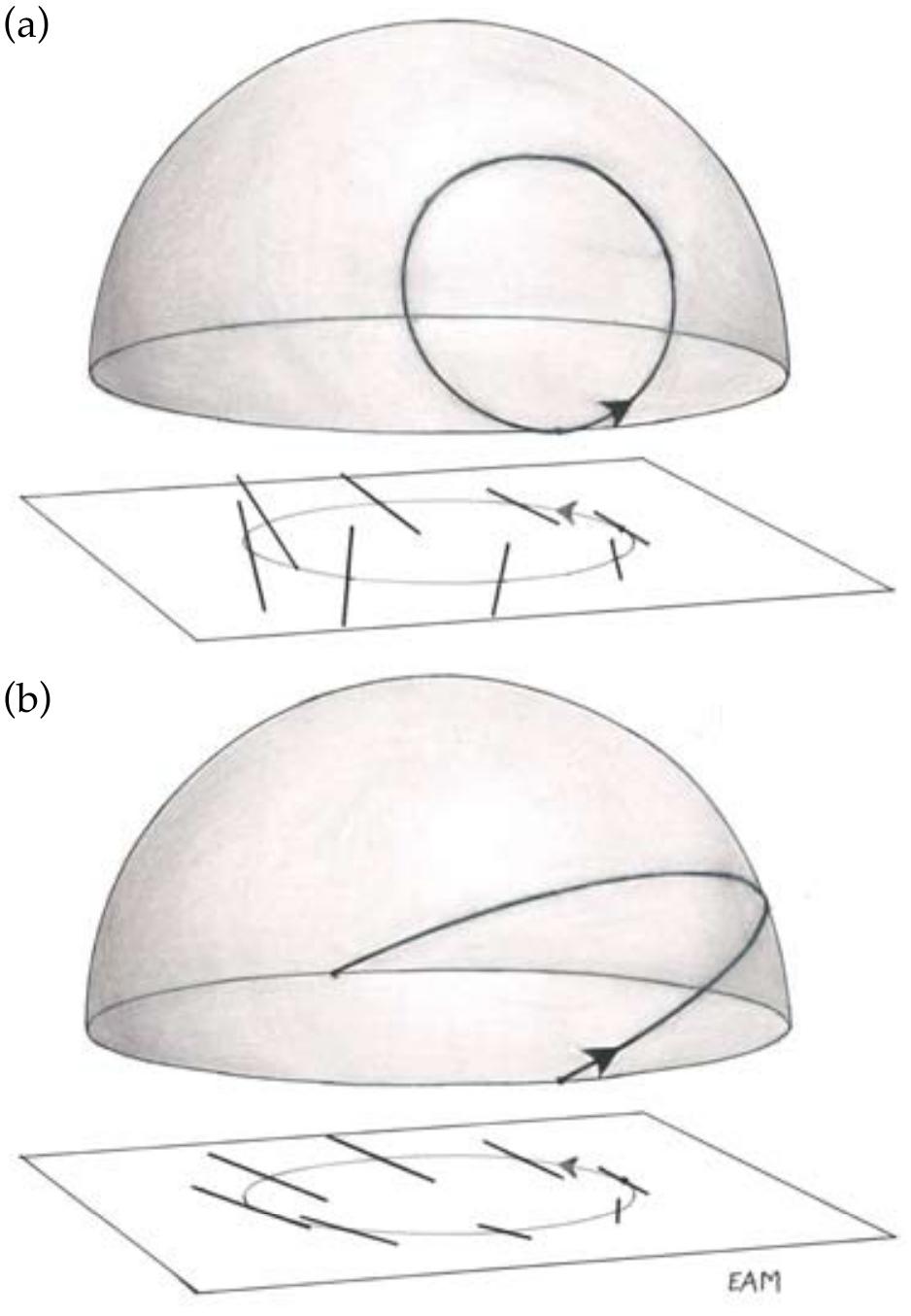}
\caption{The GSM for the three-dimensional nematic is the real projective plane $\mathbb{RP}^2.$  Any measurement around a closed loop in the sample can be mapped to a path in $\RP^2$ which belongs to one of two homotopy classes; trivial (a) and nontrivial (b). In other words, $\pi_1(\RP^2)=\Z_2.$}
\label{fig:RP2}
\end{center}
\end{figure}

Recall that any rotation in
three dimensions is an element of $SO(3)$, and one set of coordinates
on this space consists of the three Euler angles. Write the rotation matrices as ${\bf R}_{\alpha\beta\gamma} = {\bf N}_\alpha {\bf M}_\beta {\bf N}_\gamma$ where the matrices ${\bf M}_\beta$ and ${\bf N}_\gamma$ are
\begin{equation}
{\bf M}_\beta = \left[\begin{matrix} \cos\beta & 0 & \sin\beta \\ 0 & 1 & 0\\
  -\sin\beta &0 &\cos\beta\\\end{matrix}\right],
  {\bf N}_\gamma  =
  \left[\begin{matrix} \cos\gamma & \sin\gamma & 0 \\ -\sin\gamma &\cos\gamma& 0\\0 & 0 & 1\\\end{matrix}\right],
\end{equation}
and $\alpha,\gamma\in[0,2\pi)$ and $\beta\in[0,\pi]$. Away from the defect, the director is a  rotation of some fiducial direction, say
$\hat{\bf z}$, so at each point $\vec{x}$ in space there are three
angles $\alpha({\vec x}),\beta({\vec x})$, and $\gamma({\vec x})$, and
${\bf n} = {\bf R}_{\alpha\beta\gamma} {\hat{\bf z}}$.  The uniaxial
nematic state has two symmetries.  First, because it is uniaxial, rotations around ${\bf n}$ leave the system invariant (a phase made of {\sl cylinders}) and so we must identify ${\bf R}_{\alpha\beta\gamma}$ with ${\bf R}_{\alpha\beta\gamma'}$ so that a rotation around the original $\hat{z}$-axis does not change the state.  Since we may as well set $\gamma'=0$, we first see that we can parameterize all directors by only two angles and we
only need consider the subspace of $SO(3)$ represented by ${\bf
T}_{\alpha\beta} = {\bf R}_{\alpha\beta 0}$.  The group of rotations
around a single axis is called $SO(2)$ and this subspace made by ${\bf
T}_{\alpha\beta}$ is called $SO(3)/SO(2)$ -- elements of $SO(3)$ which
are identified with each other by $SO(2)$ rotations.  Observe that
this is a sphere, the space of configurations of a unit vector, appropriate for a spin in a Heisenberg magnet, and $\alpha$ and $\beta$ run over the standard azimuthal and polar angles on the sphere, respectively. 
In a nematic, ${\bf n}$ and $-{\bf n}$ are further identified; they
are related by a rotation by $\pi$ around any axis perpendicular to the director ${\bf n}$. 
Again, we can pre-rotate the $\hat{\bf z}$ direction by $\pi$ using
the diagonal matrix ${\bf P} = \hbox{diag}[1,-1,-1]$, an element of
$SO(3)$ with ${\bf P}^2=\bf 1$.  The two element group made of $\bf 1$
and $\bf P$ is called $\mathbb{Z}_2$. The nematic symmetry forces us to identify the two elements
${\bf T}_{\alpha\beta}$ and ${\bf T}_{\alpha\beta}{\bf P}={\bf
T}_{\alpha +\pi,\pi-\beta}$.  The resulting space is $SO(3)/H$, where $H$ is the group of nematic symmetries (the isotropy subgroup) which includes {\sl both} $SO(2)$ and $\mathbb{Z}_2$.  Again, the notation $SO(3)/H$ indicates that we identify two elements of $SO(3)$, ${\bf R}$ and ${\bf R}'$, if they differ by any element of the group made of $H=\{{\bf N}_\alpha,{\bf P}{\bf N}_\alpha\}$ for all $\alpha$.  Note that ${\bf P}{\bf N}_\alpha{\bf P} = {\bf N}_{-\alpha}$ and so the product of two elements of $H$, ${\bf P}^{m_1}{\bf N}_{\alpha_1}$ and ${\bf P}^{m_2}{\bf N}_{\alpha_2}$ is 
\begin{equation}\label{eq:semidirect}
{\bf P}^{m_1} {\bf N}_{\alpha_1} {\bf P}^{m_2} {\bf N}_{\alpha_2} = {\bf P}^{m_1+m_2}{\bf N}_{\alpha_2 + (-1)^{m_2}\alpha_1}.
\end{equation}
While the total number of flips just adds ($m_1 + m_2$), the rotation
angles do not -- whether they are added or subtracted depends on
$m_2$.  Thus the group addition in $H$ is not just the group
addition of the two different parts $\mathbb{Z}_2$ and $SO(2)$
separately for $m$ and $\alpha$.
$H$ is called the semidirect product of the two groups and, in this
case, is the point symmetry group $D_{\infty}$, the proper symmetries
of a right cylinder with constant, circular cross-section.  In any
event, the resulting space $SO(3)/H$ has a name, the real projective
plane $\mathbb{RP}^2$, and the identification of the spherical angles
$(\theta,\phi)\leftrightarrow (\pi-\theta,\phi+\pi)$ allows us to view
it, just as our intuition suggested, as the sphere with
diametrically opposed points identified, 
as shown in FIG.~\ref{fig:RP2}.  However, the way that the angles add \eqref{eq:semidirect} may not
have occurred to the casual observer without this analysis\footnote{Note that in our discussion of the two
dimensional nematic, we could have described the GSM as $\mathbb{S}^1$
with diameters identified and gone through the same construction.  The
resulting space is $\mathbb{RP}^1$, but its topology is {\sl
identical} to that of $\mathbb{S}^1$. $\mathbb{S}^2$ and
$\mathbb{RP}^2$ are {\sl not} topologically identical; in particular
they have different fundamental groups.}.

\begin{figure}[t]
\includegraphics[width=3.25in]{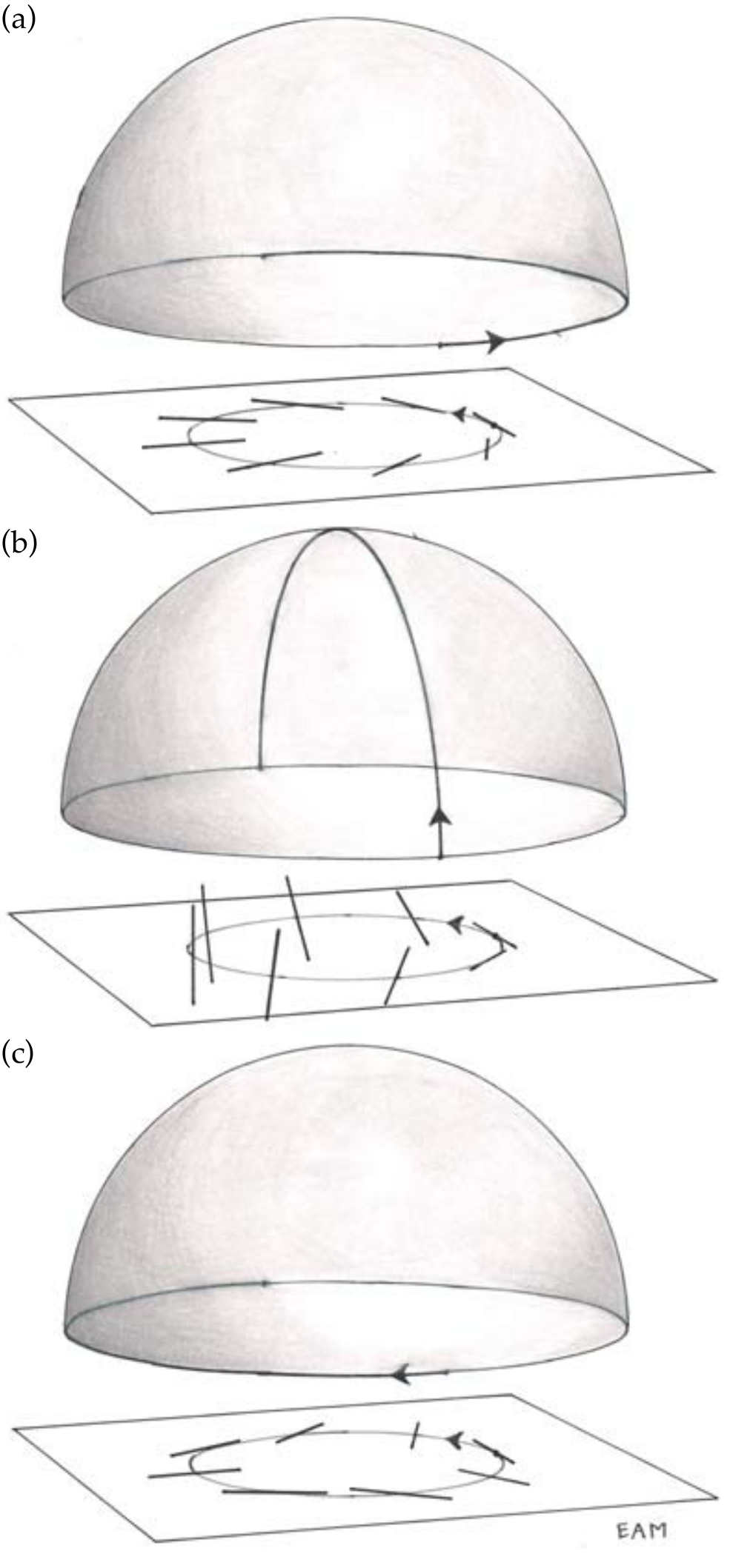}
\caption{The homotopy that takes a $+\frac{1}2$ disclination line in a three-dimensional nematic into a $-\frac{1}2$ disclination line consists of a uniform rotation of all the molecules by $\pi$ about the $\hat x$-axis.  In $\RP^2,$ this homotopy entails sliding the path from the eastern half of the equator (a) over the North Pole (b) to the western half of the equator (c).}
\label{fig:rp2_homotopy}
\end{figure}

This analysis demonstrates that based measuring circuits with the topology of
a circle in a three-dimensional nematic and the one-dimensional
defect structures they may encircle
are characterized not by $\pi_1(\mathbb{RP}^1)$ but rather by
$\pi_1(\mathbb{RP}^2)$, which turns out to be a much smaller group. 
Although when we traverse our measuring circuit in the sample the final orientation of the director field must be the same as the initial one, the final rotation ${\bf R}_{\alpha\beta\gamma}$ does not have to simply be the identity; it can be any symmetry transformation of the nematic, {\sl i.e.} any element of $H=D_{\infty}$. This subgroup consists of two disconnected pieces, the rotations ${\bf N}_{\alpha}$ and composite rotations ${\bf PN}_{\alpha}$. Homotopies can alter the precise value of the final rotation but not which connected component it is in. Consequently we see that there are two classes of loops in $\mathbb{RP}^2$. Returning to the geometric representation of $\mathbb{RP}^2$ in FIG. \ref{fig:RP2} as the sphere with antipodal points identified, we can see that these two classes can be identified with loops that start and end at the same point on the sphere and those that start and end at antipodal points.  
The presence of a defect inside our measuring loop
corresponds to the latter class of paths.  In the previous case, we
could simply count the number of brushes to determine that
$\pi_1(\mathbb{RP}^1) = \frac{1}{2}\mathbb{Z}$, the half-integers.  In
the case of a true three-dimensional director we see that the group is
actually simpler: $\pi_1(\mathbb{RP}^2)=\mathbb{Z}_2$ since any closed
path on $\mathbb{RP}^2$ can either be smoothly deformed (holding its
base point fixed) to its base point or
to a path that connects antipodes.  We will write $\mathbb{Z}_2$
multiplicatively, so that it has the two elements $\{1,-1\}$. 
Thus, no distinction 
can be made any longer between positive and negative strength defects; the planar
textures ${\bf c}_{1/2}$ and ${\bf c}_{-1/2}$ can be converted into
each other by a uniform rotation of all of the molecules by $\pi$
about some axis in the plane, $\hat{x}$ say. On the GSM this rotation
simply corresponds to transporting the loop ${\bf n}(\Gamma)$ from one
side of the equator to the other by passing it over the North Pole, as shown in FIG. \ref{fig:rp2_homotopy}.
Equally, there are no `winding 1' defects; the combination of circuits
around two disclination lines in this case is always homotopic to a
defect free texture. For instance, the combination of the two planar
textures ${\bf c}_{1/2}$ produces a loop on the GSM that goes all the
way around the equator and so can be smoothly slid up in latitude
until it shrinks to a point at the North Pole\footnote{Strictly this
particular homotopy does not preserve the base point, however, it is
easy to construct one that does; simply hold that point of the loop
fixed on the equator, slide the rest over the North Pole and shrink it
into the base point.  In this case too, $[\mathbb{S}^1,\mathbb{RP}^2]$ has the
same elements as $\pi_1(\mathbb{RP}^2)$.  In general
$[\mathbb{S}^1,\text{GSM}]$ is equal to the underlying set of the group
$\pi_1(\text{GSM})$ whenever
the latter is Abelian.}.

The above homotopy from a loop covering the equator to a constant loop
at the North Pole is actually observed in cylindrical capillaries treated to give perpendicular anchoring at the surface. The molecules rotate out of the plane so as to align along the cylinder axis in the interior -- the so called `escape in the third dimension'~\cite{cladis72,meyer73,williams74}. Of course, this escape can proceed in two directions: up or down. If, as typically happens, the molecules escape up in one section of the capillary and down in another then there will be a mismatch in between. This gives rise to a different type of defect: point defects in three dimensions, known colloquially in the field as hedgehogs~\cite{polyakov74}.

\section{Hedgehogs}
\label{sec:hedgehogs}

Now that we have been introduced to the line defect, it behooves us to meet its partner, the hedgehog.
Many features of 
the description of line defects carry over to
the point-like hedgehogs. 
In this paragraph, we outline what is to come in this section by 
sketching a few ``3-dimensional versions'' of the pictures described 
previously.  First, a measuring loop can link with a defect
line, but a loop cannot link with a point.  Just as we need a Gaussian
sphere to measure a point charge in electrostatics, we measure the topological charge of
a point defect on a measuring sphere (denoted $\mathbb{S}^2$) in our sample.  
Thus what we seek to classify are the set of maps without base point
$[\mathbb{S}^2,\mathbb{RP}^2]$ and the second homotopy group 
$\pi_2(\mathbb{RP}^2)$.  With that knowledge in hand, we can consider a 
piece of the sample with multiple point defects by imagining a 
three-dimensional ball with measuring spheres around punctures at
the hedgehogs.  This space retracts to a bouquet of spheres,
and the ``sum'' of the textures on the measuring spheres in the 
bouquet is taken to be the texture on a sphere which shrink-wraps the 
entire thing. 
Just as for line defects, this picture relates defect 
addition and a local-to-global construction.

\begin{figure*}[!t]
\includegraphics[width=6.5in]{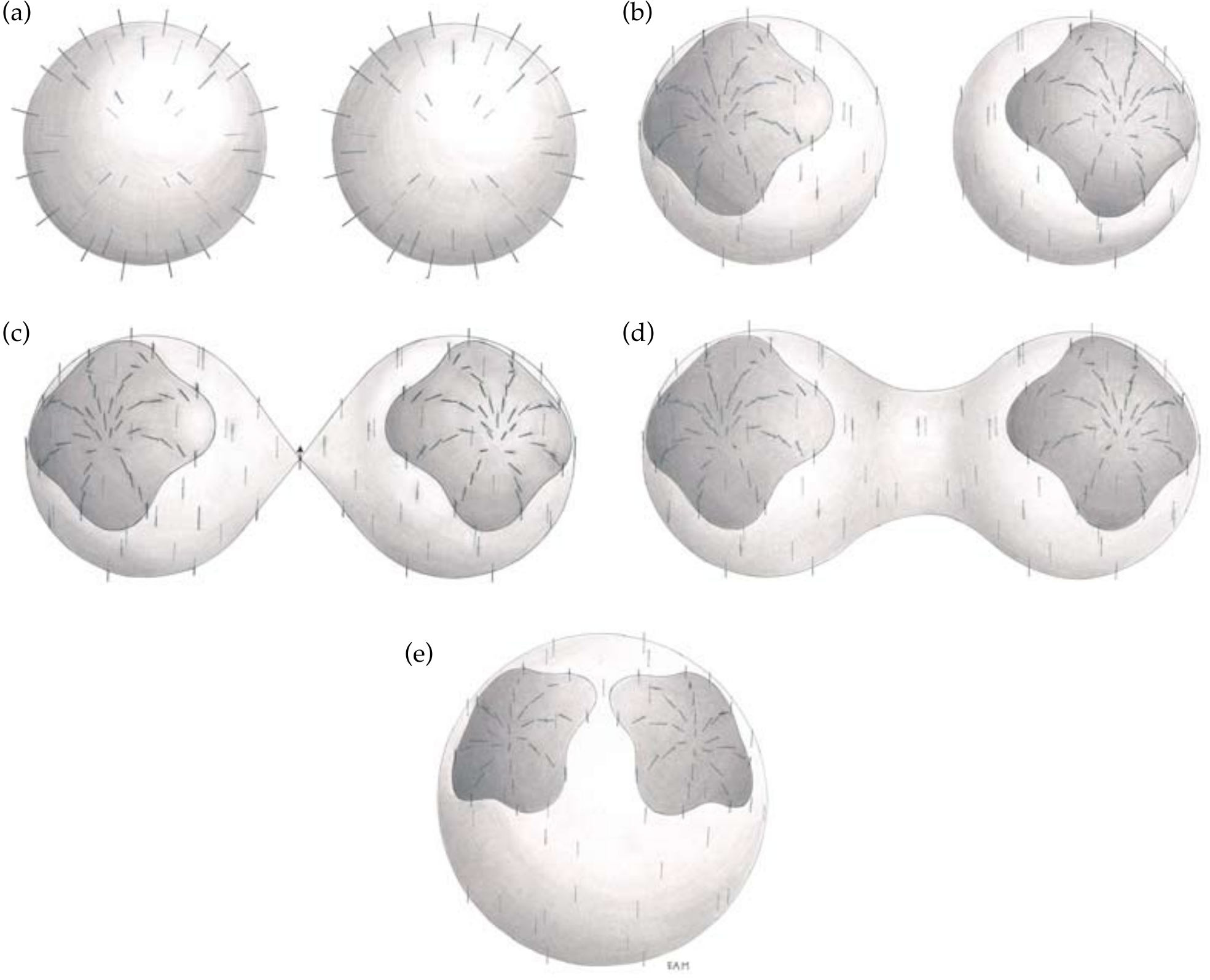}
\caption{How do you combine two hedgehog charges (a)?  The key lies in based homotopy theory.  First, we comb the texture on the measuring surface so that the defect is confined to a small island within an ocean of uniform director field, which we identify as the base point (b).  In order to compare the relative charges of two hedgehogs, their respective measuring surfaces must originate from the same base point, forming the bouquet of spheres described in the text (c).  A new measuring surface, a sphere which has been shrink-wrapped around the bouquet of spheres, now contains both islands (d) and may be deformed into a single sphere containing the sum of both hedgehogs (e).}
\label{fig:adding_hedgehogs}
\end{figure*}

Let us first study the group structure on $\pi_2(\text{GSM})$,
consisting of homotopy classes of based maps from the sphere to the GSM.
Here again, we fix a base point on both the sphere (which
corresponds to the base point in our sample) and the GSM to
define a group addition property.  To be precise, consider the
standard latitude and longitude coordinates on the globe.  We choose
the North Pole for the standard point on the sphere and some
convenient point, say $\Upsilon_0$, in the GSM. To combine elements we imagine the following sequence of homotopies,
depicted in FIG. \ref{fig:adding_hedgehogs}.
Consider first a single element
$g_1\in\pi_2(\text{GSM})$. We can smoothly deform the map by combing
out a neighborhood of the North Pole into a polar cap which maps
entirely to $\Upsilon_0$. We can keep smoothly deforming the whole map
into a smaller and smaller patch of the sphere, until we have a small
island surrounded by an ocean of points all mapping to $\Upsilon_0$. 
The texture arising on the island from a standard $+1$ hedgehog is a skyrmion~\cite{rossler06,muhlbauer09} and the director rotates from $\Upsilon_0$ at the coast all the way to the antipodal direction $\bar{\Upsilon}_0$ in the interior, covering every orientation in between. 
This new map is homotopic to the original map which corresponded to
$g_1$ and so, from the point of view of $\pi_2(\text{GSM})$, the new
map is the same element $g_1$ -- remember, the rate at which we
traverse the GSM is not important, only the places visited.   We can
make the same smooth maneuver on a second element
$g_2\in\pi_2(\text{GSM})$, making a second island in the same ocean.
To combine the two elements, we simply put island one ($g_1$) and
island two ($g_2$) in the same ocean without overlapping the islands.
This works because the boundaries of these islands are all mapped to
$\Upsilon_0$.  We can now deform this new map smoothly as we see fit,
filling the ocean back up with land.  In doing so we have combined the
two elements.  Since we can move the islands around before filling in the ocean, there is no way to order the group addition and so $\pi_2(\text{GSM})$ and it follows that $\pi_2$ is necessarily Abelian.  As in the case of $\pi_1$, the identity element is
again the equivalence class of the uniform texture that maps the whole sphere to $\Upsilon_0$.  

\begin{figure}[!h]
\begin{center}
\includegraphics[width=3.25in]{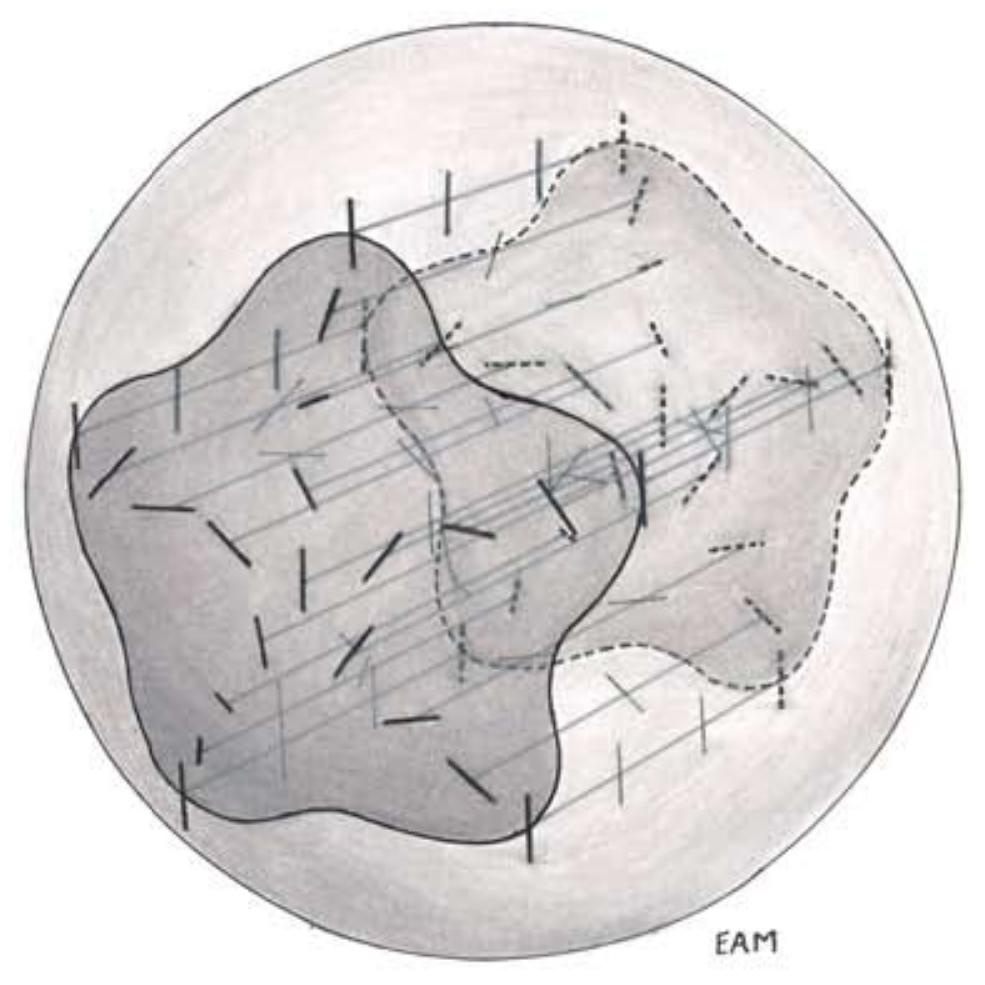}
\caption{The inverse of an element consists of taking the mirror image of its island and placing this new island on the sphere.  The sum of these two elements must add so that the sphere contains no defects.  In the interior of the sphere, the texture is constant along chords connecting identical points on the original island and its mirror.  Because no defects are introduced, the entire texture on the sphere may now be smoothly deformed to the base point.}
\label{fig:inverse}
\end{center}
\end{figure}

Finally, the inverse of any element can be found by deforming the original map
into an island and then flipping the island over on the sphere.  Why
does this give an inverse? We can connect each point on the Western
hemisphere to its reflection through the plane including the prime
meridian in the Eastern hemisphere.  We can choose the value of the
map on each of these lines inside the two sphere to take the value at
the endpoints -- identical by construction.  It follows that we can
fill the region inside the sphere with a smooth texture and there can be no defects
inside.  Thus the island and its mirror add to zero, shown in FIG. \ref{fig:inverse}. 
Alternatively, we can take the map
$f(\theta,\phi)\mapsto \text{GSM}$ and create the inverse
$g(\theta,\phi)=f(\theta,-\phi)$.  

Let's relate the group structure of $\pi_2(\text{GSM})$ to the addition of defects in terms of 
measuring surfaces: imagine (as in FIG. \ref{fig:adding_hedgehogs}) that we measure two different regions, each with
a sphere.  As for loops, in order to add the sphere measurements, they must be in a bouquet
configuration.  We can achieve this by
attaching a tubelike tether to each sphere ending at our base point.
Beware! The charges assigned to hedgehogs (actually, assigned to measuring spheres) will depend on the choice of tethers, 
particularly if there are disclinations in the system. 
The sum of the textures on these spheres is equivalent to the texture on a large sphere which has been deformed so that it snugly wraps around all of the original spheres.
We encourage readers to relate this to the earlier picture of islands on 
a single sphere -- each sphere in the bouquet becomes an island, and the base
point of the bouquet is blown up to form the ocean.   If {\sl no disclination lines are present}, {\sl i.e.} $\pi_1(\text{GSM})=\{1\}$, then the tethers can always be unentangled since you ``can't lasso a basketball'' \cite{colemanbook}.

Though we have a definition and interpretation of $\pi_2(\mathbb{RP}^2)$,
we haven't computed it yet.  In order to do so, we 
generalize the idea of counting brushes in the two-dimensional nematic.  Each brush is a place where the director points in one of two directions -- parallel to the polarizer or analyzer -- and the charge is one quarter the number of brushes.  Because of the cross-polarizers, we are forced to count both the polarizer and analyzer directions.  However, were we given an explicit map we could, instead, just count the number of times the map from $\mathbb{S}^1$ went to any single element of $\mathbb{RP}^1$ (with sign, as in Eq. \eqref{eq:winding}), an integer known as the {\sl degree} of the map.  Then the charge of the defect would be half the degree.  We could try to do the same thing for maps from $\mathbb{S}^2$ to $\mathbb{RP}^2$, but the ambiguity of $\bf n$ and $-\bf n$ makes this problematic\footnote{The reader may worry that the same problem plagues us for $\pi_1(\mathbb{RP}^1)$.  It does not: $\mathbb{RP}^1$ is orientable where $\mathbb{RP}^2$ is not, which is the real issue here.}. 
Each point of $\mathbb{RP}^2$ can be identified with two antipodal
points on $\mathbb{S}^2$:  we can start at the base point $\Upsilon_0$
and choose one of the two equivalent points on $\mathbb{S}^2$.
Given a choice, we can do this for all points in a small neighborhood
consistently so that we locally generate a new smooth map to
$\mathbb{S}^2$. The only hitch might be that when taking a long path
around $\mathbb{RP}^2$ we may find that we are hung up on a
non-contractible path in the GSM when we return to the same
point\footnote{This is the distinction between the identity and $-1$ in
$\pi_1(\mathbb{RP}^2)$: when paths in $\mathbb{RP}^2$ are lifted to
paths in $\mathbb{S}^2$, paths in the equivalence class of the identity are closed, and paths
in the class $-1$ are not.}.
However, this is {\sl impossible} with maps from $\mathbb{S}^2$.  Fortunately, maps from $\mathbb{S}^2$ to any space $X$ have a special property owing to the topology of the sphere.  Consider a closed loop $\Gamma$ on $\mathbb{S}^2$ and the loop it produces on $X$ and watch how the latter changes as we smoothly deform $\Gamma$.  Because $\Gamma$ can be shrunk to a point on $\mathbb{S}^2$, 
its image on $X$ can also be shrunk to a point and to one value in $X$.  For the case of $\mathbb{RP}^2$ this is very convenient: the image of any loop $\Gamma$ on $\mathbb{S}^2$ must be homotopic to the identity map on $\mathbb{RP}^2$, so we never get hung up. Once again, we cannot lasso a basketball and so  
we can always ``lift'' the map from $\mathbb{RP}^2$ to $\mathbb{S}^2$
globally and turn it into a map from $\mathbb{S}^2$ to $\mathbb{S}^2$.
Therefore, after fixing the lift of
$\Upsilon_0$ (for definitiveness, we'll take it to lie in the Northern
hemisphere), $\pi_2(\mathbb{RP}^2)$ is the same as
$\pi_2(\mathbb{S}^2)$, which we can calculate.  

Each map gives us a unit vector $\bf{n}(\theta,\phi)$ at each point of the
measuring sphere.  Like the winding number, there is an integral that
measures the number of times our map wraps around the sphere, also
known as the {\sl degree} or {\sl hedgehog charge}
\begin{equation}
d=\frac{1}{4\pi} \int_{\mathbb{S}^2} d\theta d\phi\, {\bf n}\cdot\left[\partial_\theta{\bf n}\times\partial_\phi{\bf n}\right].
\label{eq:degree}
\end{equation}
Remarkably, it is an integer: the integrand is just the Jacobian of the
map from $\mathbb{S}^2$ to $\mathbb{S}^2$ and so it counts the area
swept out on the target sphere.  Dividing by $4\pi$ simply gives us
the number of times we visit each point on the target.  Moreover,
because we do not take the absolute value of the Jacobian, we measure
positive and negative area and thus will properly count the analog of
increasing and decreasing as needed in \eqref{eq:winding} for the
winding number.  See, for
instance, \cite{Kamien02} for the derivation of this Jacobian and its
connection to Gaussian curvature. Therefore, we see that maps from $\mathbb{S}^2$ to $\mathbb{S}^2$ are classified by a degree and that this degree can be any integer, positive or negative. Thus $\pi_2(\mathbb{RP}^2)=\pi_2(\mathbb{S}^2)=\mathbb{Z}$ and there are an infinite number of topologically distinct point defects in nematic liquid crystals. Representatives of each homotopy class are given by the maps 
\begin{equation} 
{\bf n}_d(\theta,\phi) = \bigl[ \sin(\theta)\cos(d\phi), \sin(\theta)\sin(d\phi), \cos(\theta) \bigr], 
\label{eq:hedgehogs}
\end{equation} 
which exhibit a $d$-fold winding on the equator and may be thought of
like the textures ${\bf c}_m$ of equation \eqref{eq:schlieren} as
providing an idealized description of the director field near a hedgehog. 

There remains just one point to straighten out: the question of the
choice of lift. This will also tell us about the difference between
the free homotopy classes $[\mathbb{S}^2,\mathbb{RP}^2]$ and
the second homotopy group $\pi_2(\mathbb{RP}^2)$.  Suppose we had
instead chosen to lift the base point $\Upsilon_0$ into the Southern
hemisphere of $\mathbb{S}^2$. What would be different?
We would still be able to identify each element of $\pi_2(\mathbb{RP}^2)$ with one in $\pi_2(\mathbb{S}^2)$ and define a degree $d$ through the formula \eqref{eq:degree}. However, when we would have used the vector ${\bf n}(\theta,\phi)$ to specify our texture we would now use $-{\bf n}(\theta,\phi)$. Since equation \eqref{eq:degree} is odd in ${\bf n}$ the degree we now measure would be $-d$ instead of $d$.
This is
not to say that the textures labelled by $d$ and $-d$ are the same;
the identification we made between $\pi_2(\mathbb{RP}^2)$ and
$\pi_2(\mathbb{S}^2)$ is one-to-one, it's just that there are two
identifications that we can make, differing in whether we lift the
base point to the Northern or Southern hemisphere, and these are not
the same\footnote{The reader might be concerned that the same
ambiguity arises for $\pi_1(\mathbb{RP}^1)$. However the two choices
of lift from $\mathbb{RP}^1$ to $\mathbb{S}^1$ differ by a $\pi$
rotation, i.e., to replacing $f_t(\phi)$ with $f_t(\phi)+\pi$ in
equation \eqref{eq:winding}, which does not change the winding number
$w$. In this case both lifts yield the {\sl same} identification of
$\pi_1(\mathbb{RP}^1)$ with $\pi_1(\mathbb{S}^1)$.}. So long as we
make our choice of lift consistently, hedgehogs with charge $d$ and
$-d$ are distinct, but it is up to us which sign we associate to which
hedgehog\footnote{The prevailing convention in the literature is to
consider the purely radial texture, equation \eqref{eq:hedgehogs} with
$d=1$, to have positive charge, the choice of lift being that the
director field points outwards.  The base point on the
measuring sphere would be the North Pole and the base point in
$\mathbb{RP}^2$ is the vertical direction, which is then chosen to lift to
the North Pole in $\mathbb{S}^2$.}.  

However, under free homotopy maps from the sphere to $\mathbb{RP}^2$ 
labeled by $d$ and $-d$ do become equivalent -- take the texture of \eqref{eq:hedgehogs} and rotate uniformly by $\pi$ around the $\hat y$-axis, then take $\bf n$ to $-\bf n$ to change $d$ to $-d$.  It follows that
$[\mathbb{S}^2,\mathbb{RP}^2]$ is  the set of {\sl
nonnegative} integers, which is not the  set of elements in
$\pi_2(\mathbb{RP}^2)$!  
Thus even with the same spherical measuring surface, the classification of 
hedgehog defects can be different depending on how we make the lift from $\mathbb{RP}^2$ and $\mathbb{S}^2$.  As long as we don't try to extend the local measurements to a global texture, the ``local-to-global problem'', we need not be careful.  However, if we specify free homotopy classes on spheres about defects and then attempt to fill the remaining space, we run into ambiguities -- trouble!
Were we to probe each hedgehog individually by calculating the degree
on a small sphere around it,  we could
only measure its charge up to a sign and we are assigning it to one of the maps $[\mathbb{S}^2,\mathbb{RP}^2]$.  The ease with which
we do this extracts a penalty -- we can no longer add defects since we have lost the group structure of $\pi_2(\mathbb{RP}^2)$ without the use of a base point.
There is no free lunch, regrettably.
For instance, a pair of spheres with
$|d|=1$ on both might induce either $|d|=$0 or 2 -- we can't know the
relative signs unless we have a common reference coming from a fixed
base point.  

To understand the global structure it is especially illustrative to view the sample as a bouquet of spheres attached at a single point, the base point of the sample, that maps to the base point of the GSM.  If we have $k$ point defects then the texture becomes a map from  a bouquet of $k$ spheres to $\mathbb{RP}^2$.  Were we to measure the ``charge'' on each sphere using the base points, we would end up with a $k$-tuple of elements of $\pi_2(\mathbb{RP}^2)$ labelled by their degree, $(d_1,d_2,\ldots,d_k)$.  There is always a global ambiguity when we lift to $\mathbb{S}^2$ since the GSM base point can be taken to lie in either the northern or southern hemisphere.  If we now choose to consider the free homotopy classes of samples with $k$ defects, we identify $(d_1,d_2,\ldots,d_k)$ with $(-d_1,-d_2,\ldots,-d_k)$, {\sl not} $(\pm d_1,\pm d_2,\ldots,\pm d_k)$
In other words, it is possible to go from a based measurement of a single defect to
a based measurement of a collection of defects to an unbased measurement of many defects.  It
is not, however, possible to make an unbased measurement of many defects from unbased measurements
of single defects.  This is the crux of the local-to-global problem and is why local information is not always enough
to understand the topology of the whole texture.

\section{Disclination loops}
\label{sec:loops}

We have now described disclination lines and hedgehogs separately, but
in any typical texture both will be present simultaneously and we are
forced to think about how they interact with each other.  When we discussed defect lines previously, we merely generalized point defects into a third direction and classified them by measuring circuits which ensnare the defect lines.  However, these measurements could not determine whether or not the 
disclination itself is a loop or a knot, or linked with other disclinations.
Indeed, a great part of the 
appeal of colloidal
systems is precisely that they provide a natural setting for
illustrating the topological interplay between disclinations and
hedgehogs and enable the braiding and tangling of the disclination lines \cite{tkalec11}. 
If the colloid is treated to promote radial anchoring of
the molecules at its surface then it will appear like a unit strength
hedgehog and this topological charge will have to be compensated by a
companion defect in the liquid crystal so as to satisfy the global
boundary conditions.  
Another way of saying this 
is that the homotopy class of the texture on a measuring sphere which 
surrounds just the colloid differs from the one on a much larger measuring
sphere on which, say, the texture is determined by the global
boundary conditions. This implies the existence of another defect set.  
One way of achieving this balance is to place an opposite strength point defect close to the colloid so that together they form a dipole pair~\cite{poulin97}. Multiple dipolar colloids interact with each other elastically through the distortions they produce in the liquid crystal and can as a consequence assemble into chains or two-dimensional colloidal crystals~\cite{musevic06,skarabot07}. 

A separate possibility is to have the charge of the colloid
compensated by a disclination loop, encircling the particle in a
`Saturn ring' configuration~\cite{terentjev95}. These too interact
elastically and aggregate to form chains and
crystals~\cite{skarabot08}. However, they also allow for a variety of
intriguing entangled structures where a single disclination loop wraps
around two or more colloids, balancing their collective
charge~\cite{guzman03,ravnik07,ravnik09}. This naturally begs the
question of how to describe the topological properties of such
disclination loops, which may be probed both as line and
point defects. 
Although this makes sense from the perspective of
conserving total charge, hedgehog charge, it turns out, is not a
homotopy invariant of a the full measuring surface of a disclination loop -- a torus. 

\subsection{Measuring with Spheres}

\begin{figure*}[!h]
\begin{center}
\includegraphics[width=6.5in]{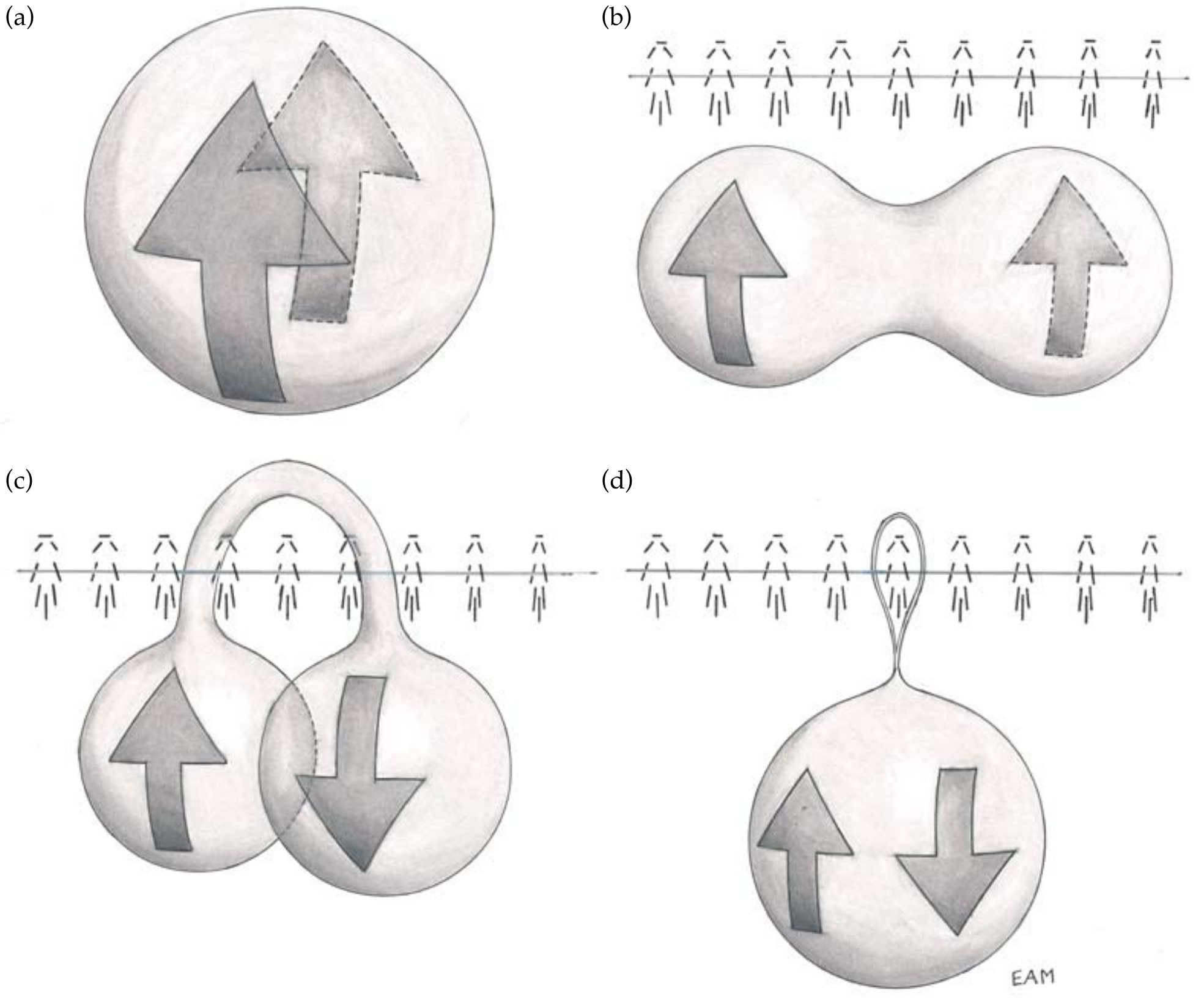}
\caption{A sphere with a plus and minus island has zero net hedgehog charge (a). The two islands can be separated to create a plus ($S_+$) and minus ($S_-$) pair (b). As the minus hedgehog is carried around a disclination line its orientation reverses as the texture on the tether between the two dumbbells connects antipodal points of $\mathbb{RP}^2$ and the natural final measuring circuit ($S_-'$) records the opposite hedgehog charge (c). If the two hedgehogs are now recombined and the tether is ignored their total charge will be $\pm 2$! (d) 
}
\label{fig:plus_minus_1}
\end{center}
\end{figure*}

Consider a based measuring sphere in our sample which doesn't surround
any defect.  The nematic texture can be deformed so that the sphere 
has one island with a +1 charge and a mirror island with charge -1.  
As shown in FIG. \ref{fig:plus_minus_1},
we can imagine distorting the sphere into a dumbbell shape with one island on each
end.  As we pinch off the tether connecting the two ends, each island comes to reside on its own sphere, $S_+$ and $S_-$, surrounding a plus and minus 
hedgehog, respectively, which we realize were necessarily created during the process.  
Suppose that elsewhere in the sample there is a 1/2 disclination line.
What happens if we drag the -1 hedgehog around that 
disclination?
More precisely, we keep the texture fixed on $S_+$ so that the degree integral is unchanged but begin
deforming $S_-$ so that it wraps once around the disclination
line and returns to its original position, leaving a tail tethered around
the disclination line.  Now, the most natural sphere around the -1 hedgehog 
is one that doesn't have the tether, $S_-'$, and is the one which we would likely use to measure
the charge locally.  What's the relation
between the textures on $S_-$ and $S_-'$?
It can be difficult to visualize this operation, which
involves a homotopy of a three-dimensional line field, but if we
return to the original undeformed texture, the texture on $S_-'$ is in
fact equivalent to the one on $S_-$ except that we then have
attached a loop to it which goes around 
the disclination line the ``opposite direction'' to the base point.  
Roughly speaking $S_-$ and $S_-'$ clasp the
hedgehog from different sides of the line defect. 
Let us call this set of operations maneuver $\mathbb{X}$. 
Such a tether picks up the $\pi$ rotation of the
disclination and we are forced to change $\bf n$
to $-\bf n$ on most of the sphere, in particular over the island, when we lift the texture on $S_-'$ from
$\mathbb{RP}^2$ to $\mathbb{S}^2$. Thus when we calculate the degree on $S_-'$ via integration, $d$ will go to
$-d$!  We have reversed the degree of the map on the sphere around the
hedgehog by ``moving it around a disclination.''  Note that we had to
carefully compare {\sl two} distinct measuring spheres around the hedgehog in question
in order to make sense of this.
We started with total charge 0 around the two hedgehogs and 
ended with total charge $\pm 2$ because we changed the measuring surfaces -- just as the total charge
in a Gaussian measuring box cannot change, neither can the charge in fixed measuring circuit.

We may also interpret maneuver $\mathbb{X}$ from the point of view of
``islands on the globe.''  Imagine a disclination loop that
approaches the globe.  We may poke an island on the sphere through
this loop, and while performing this process the loop leaves an image
in the shape of an atoll in the $\Upsilon_0$-ocean.  Though the value in $\mathbb{RP}^2$
can be the same inside and outside the atoll, when we lift to
$\mathbb{S}^2$ we must change the sign of $\bf n$ when we cross the
atoll, as this is equivalent to wrapping around the disclination loop.  
If the atoll is only surrounding a region of open ocean, this
is not a problem -- we can shrink it to a point and effectively make
the whole region of ``wrong'' ocean of $\bar\Upsilon_0$ disappear.
But given an island carrying a nontrivial hedgehog charge, when we lift to
$\mathbb{S}^2$ we will be forced to match the island's coastline to a
lagoon of $\bar\Upsilon_0$.  No problem, just lift the island to $-\bf
n$ instead of $\bf n$.  Is there a singularity of the texture on the
atoll?  No!   For concreteness take $\Upsilon_0$ to be the North Pole
and so $\bar\Upsilon_0$ is the South Pole.  Topologically, the atoll
is an annulus.  The inner ring of the annulus points South and the
outer ring points North.  It is no problem to have the director
smoothly rotate along the radius of the annulus from South to North.
This texture will not contribute to the degree.  By stretching the
measuring surface this becomes precisely the picture in FIG.
\ref{fig:plus_minus_1} where the atoll becomes the tube-like tether.
Again, note that this action of $\pi_1$ on $\pi_2$ preserved the
island's class in the set of unbased maps
$[\mathbb{S}^2,\mathbb{RP}^2]$, but {\sl not} in
$\pi_2(\mathbb{RP}^2)$. 

\subsection{Measuring with Tori}
\label{subsec:measure_tori}

\begin{figure}[!h]
\begin{center}
\includegraphics[width=3.25in]{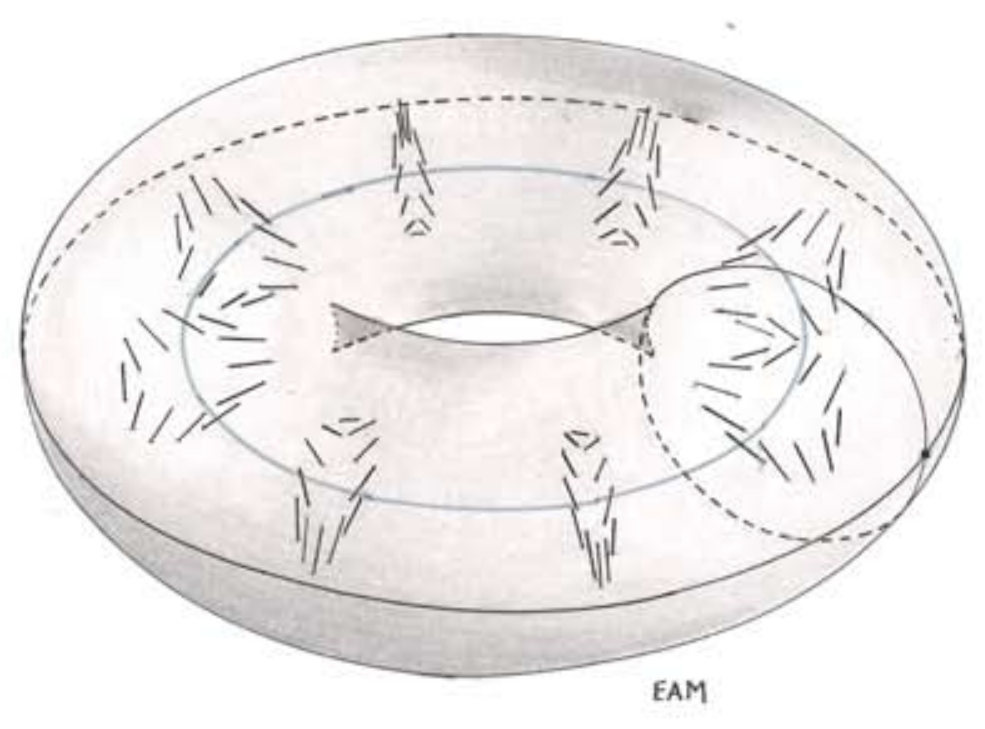}
\caption{The measuring surface $\T^2$ has two cycles, either of which may contain nontrivial winding. For the unlinked disclination loop shown the winding is non-trivial around the meridian and trivial around the equator.}
\label{fig:tori}
\end{center}
\end{figure}

Hedgehog charge is a homotopy invariant of a sphere, the natural
measuring surface for a point defect. But the natural measuring
surface for a disclination loop is not a sphere
but a torus, $\mathbb{T}^2$, a thin tube sheathing the singular line.
The classification of disclination loops can therefore be based on the homotopy
invariants of the texture on this torus, that is on maps from
$\mathbb{T}^2$ to $\mathbb{RP}^2$.  There are multiple measures of these maps, two corresponding
to
elements of $\pi_1(\mathbb{RP}^2)$
measured on the cycles of the torus and a more refined quantity, a
global defect index~\cite{janich87,nakanishi88,bechluft-sachs99} which captures some aspect
of hedgehog charge. 
A torus surrounding a simple, circular
disclination loop, shown in FIG. \ref{fig:tori}.a,
has two cycles:
a meridional loop that goes around the disclination line which always
measures the non-trivial element of $\pi_1(\mathbb{RP}^2)$ and the
loop that follows the contour of the disclination line along the
longitude of the torus records another element of
$\pi_1(\mathbb{RP}^2)$\footnote{The reader may worry about which
longitudinal path to follow if the disclination loop has a more
complicated shape or is knotted. A canonical choice is provided by any
path that has zero linking number with the defect loop
itself~\cite{janich87}.}. If this is also non-trivial then it means
that our disclination loop itself goes around another defect -- linked
loops! -- while if it is the trivial element then our disclination
loop is either isolated, or linked an even number of times. For now 
let us assume that this element is trivial and that our disclination 
loop is unlinked.

\begin{figure}[!h]
\begin{center}
\includegraphics[width=3.25in]{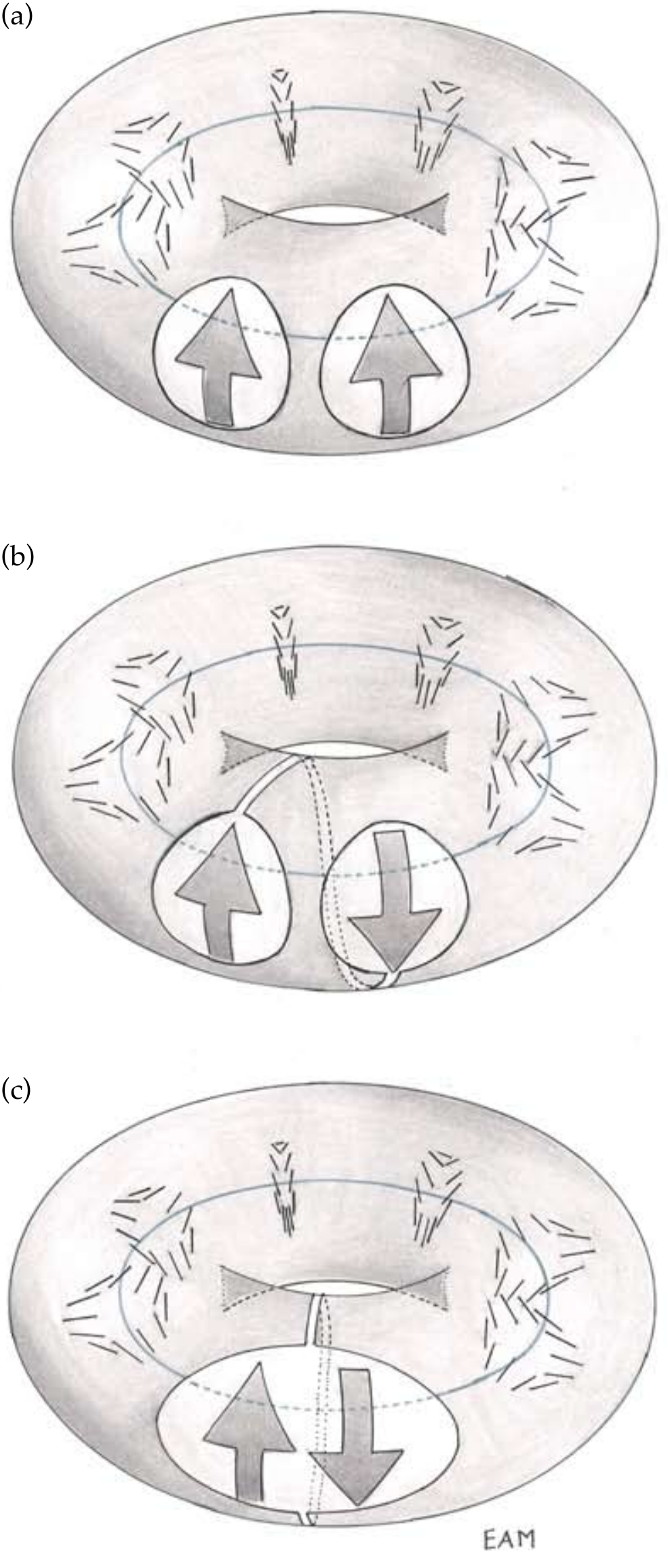}
\caption{Two $+1$ islands on a torus surrounding a disclination loop, representing the addition of two hedgehogs (a). Moving one of the islands around the meridian of the torus reverses its orientation, turning it into a charge $-1$ island (b). If the two islands are then merged they carry zero net charge, illustrating the homotopy between disclination loops with hedgehog charge $n$ and $n\pm 2$ (c).}
\label{fig:islands_on_a_torus}
\end{center}
\end{figure}

What is the homotopy classification of this subset of textures on a
torus?  One na\"ive guess would be that we would have an additional
choice of integer $n$, which we might get by repeatedly merging
hedgehogs into the line defect, in a way analogous to our way of
merging two hedgehogs together (via a bouquet construction).  
However, we can show that adding two hedgehogs to our disclination is
the same as not adding any by using maneuver $\mathbb{X}$, as in FIG. \ref{fig:islands_on_a_torus}.  Let us have a
texture on the torus called $f$; we will now show that $f'$ arising
from adding two $+1$ islands to the torus
by merging in a sphere that carries $+2$ hedgehog charge is homotopic
to $f$.   
Move one of the $+1$ islands on $f'$ on a loop
around the meridian.  As we pull the island around a circle of longitude it accumulates a concentric series of borders
around the island which rotate the coastline by $\pi$.  As before this is the action of 
$-1\in\pi_1(\mathbb{RP}^2)$ on $+1\in\pi_2(\mathbb{RP}^2)$.  This island is 
now homotopic to $-1\in\pi_2(\mathbb{RP}^2)$, as we argued before.  Now we have a $\pm1$ pair of 
islands on the torus which we may cancel against each other just as we
could for a pair of such islands on a sphere.  This
results in a texture on the torus which is homotopic to the one we started 
with.  From another point of view, performing maneuver $\mathbb{X}$ does not require any discontinuous changes on the torus -- the hedgehog moves through the handle without piercing the measuring surface.  As a result, we can smoothly transform the texture on $\mathbb{T}^2$ while changing the hedgehog charge of the surrounding space by $\pm 2$.
Hence, we can classify tori surrounding disclination loops as either carrying {\sl
even} or {\sl odd} hedgehog charge.   The reader might ask how we can
be sure that there are no moves that can change this charge by $\pm 1$, an issue that we will explain in the next section.   

In what follows we will state some interesting results that fill out
most of the rest of the story about disclination loops.
The complete classification of textures on tori labels
textures with one of the four pairs $(a,b)$ where $a,b=\pm1$ are the 
homotopy classes on the meridian and longitude cycles, respectively.  
For the three cases
where $a$ and $b$ are not both $+1$, it turns out 
there are two subclasses of textures within those that are labeled
with $(a,b)$, which correspond to even and odd numbers of +1 islands
on the torus, just as for the case $(-1,+1)$ we
explained above.  
When $a$ and $b$ are both $+1$, there are an infinite
number of classes -- without a 
disclination loop in the torus, we cannot cancel out pairs of 
islands anymore\footnote{Free and based homotopy
classes have the same classification for the first three classes where
$a,b$ are not both trivial, since multiplying by $-1$ doesn't change
evenness or oddness of the hedgehog charge on the islands.  For the
$(+1,+1)$ case, the based classification is all
integers, and the free classification is only nonnegative integers,
just as for textures on a sphere.}.  We therefore can add a 
subscript either in  $\{e,o\}$ or $\mathbb{Z}$ to this
ordered pair to complete the classification.

This peculiar set of homotopy classes has some interesting additional
structure: we may do much more with tori than just merge them with
spheres, as we did above. We present below simply one example, which
corresponds to merging two disclination loops ({\sl i.e.} tori with $a=-1$)
``side-to-side''\cite{janich87,bechluft-sachs99}.  Unfortunately, a precise explication of 
the other group
which emerges when we merge two unlinked tori ({\sl i.e.} with $b=1$)
``top-to-bottom'' \cite{nakanishi88} is just barely outside the scope
of this paper.
Given two tori with textures such that $a=-1$ on both, we cut them
along a meridional circle, resulting in two cylinders and then reglue
them so that we have a single torus.  
This results in a $\mathbb{Z}_4$ group structure on the homotopy classes of
such textures on a torus, where the mapping is $[0]=(-1,1)_e$,
$[1]=(-1,-1)_{o}$, $[2]=(-1,1)_o$, and $[3]=(-1,-1)_{e}$, where
we write $\mathbb{Z}_4$ additively so that $[m]+[n]=[m+n\mod4]$.
It's natural for $(-1,1)_e$ to be the identity element; it is both unlinked and may contain no hedgehogs, which makes adding it a bit like adding a ``constant'' segment of disclination line.

We summarize these and the results of the last three sections in Table
\ref{tab:class}.

\begin{table*}
\begin{tabular}{l|c|c|l}
& & Measuring&  \\
Physical system &GSM &Circuit&Classification / Notes\\
\hline
Point defects in Schlieren texture & $\mathbb{RP}^1$ & $\mathbb{S}^1$ &
$\frac{1}{2}\mathbb{Z}$, when based, can be added \ref{subsec:2d}\\ [3pt]
Disclination lines in 3D nematics & $\mathbb{RP}^2$ & $\mathbb{S}^1$ &
$\mathbb{Z}_2$, when based, can be multiplied \ref{subsec:3dlines} \\[3pt]
Point defects in 3D nematics & $\mathbb{RP}^2$ & $\mathbb{S}^2$ &
$\mathbb{N}$ when free, $\mathbb{Z}$ and may be added when based
\ref{sec:hedgehogs}. For based textures, \\&&& attaching a tether which runs over
$-1\in\pi_1(\mathbb{RP}^2)$ takes $d$ to $-d$ \ref{sec:loops}.\\[3pt]
Disclination loops in 3D nematics & $\mathbb{RP}^2$ & $\mathbb{T}^2\cong$ & 
$(a,b)_p$, where $a,b=\pm1$ and $p=\text{even}, \, \text{odd}$ if $a,b$ are not both 1,
if \\&&$\mathbb{S}^1\times\mathbb{S}^1$& $a=b=1$, then $p\in\mathbb{Z}$ if based, or $\mathbb{N}$ if free
\ref{sec:loops}.
\end{tabular}
\label{tab:class}
\caption{\hsize=6.5truein A summary of classifications of textures on various measuring
surfaces, their physical relevance, and how they may be combined. If
free or based is not specified, the underlying sets for the
classifications are the same.}
\end{table*}

\section{Biaxial nematics and the odd hedgehog}
\label{sec:biaxial}

In the last section, we argued that by a smooth deformation, a map from $\mathbb{T}^2$ to $\mathbb{RP}^2$ could absorb hedgehog charge in pairs so that the hedgehog charge could only be even or odd.  
Here, we probe this further and use the insight provided by decorating the uniaxial textures with a small amount of biaxial order. This biaxial point of view both highlights the underlying topology of the uniaxial phase and clarifies the way in which the even and odd classes of uniaxial disclination loops are distinct. 

Recall in our discussion of the nematic we discovered, via some matrix algebra, that the ground state manifold was $\mathbb{RP}^2$.  This was appropriate for a {\sl uniaxial} nematic, which had rotational symmetry around its long axis -- the rotational symmetry responsible for the $SO(2)$ factor in $H$.  A biaxial nematic has a lower symmetry, that of a brick or rectangular cuboid with three unequal lengths.  Not just a mathematical construct, biaxial liquid crystalline phases have been known for many decades.  In the past few years, discovery of thermotropic biaxial phases has renewed interest in their defect structures. Moreover, chirality and biaxiality are intimately connected \cite{harris99,priest}, and studies of blue phases \cite{grebel83,wright,dupuis05} often utilize a biaxial description. 

Returning to the notation and discussion in Section
\ref{subsec:3dlines}, we start with the biaxial molecule with long axis along $\hat z$ and a second axis along $\hat x$ (the third axis is along $\hat y$ and all three are distinguishable -- ``triaxial nematic'' might be more apt). The symmetry of the brick involves only three discrete rotations of $\pi$ around each of $\hat x$, $\hat y$, and $\hat z$. Ignoring those symmetries for the moment, we see that the original rotation matrices ${\bf R}_{\alpha\beta\gamma}$ represent an arbitrary rotation of the brick.  With the symmetries, we must now identify $\gamma$ with $\gamma\pm \pi$, so that the isotropy subgroup $H_b$ becomes
\begin{equation}
H_b = \left\{ \mathbf{1},{\bf P},{\bf N}_\pi,{\bf P}{\bf N}_\pi \right\},
\end{equation}
where the addition of elements (multiplication of the matrices) is as before.  Note that ${\bf
P}{\bf N}_\pi = {\bf M}_\pi$, so that a rotation around $\hat z$ of
$\pi$ followed by a similar rotation about $\hat x$ yields a $\pi$
rotation around  $\hat y$.  For all three of these elements, we are
never sure whether these rotations are by $\pi$ or $-\pi$.  Of course,
the reader might think that these lead to the same group elements,
which they do.  However, a {\sl loop} in $SO(3)$ that starts at $0$
rotation and ends at a $2\pi$ rotation is {\sl not} contractible and
leads to a non-trivial element of $\pi_1[SO(3)/H_b]$.  This fact, one
might recall, is often demonstrated in a class on quantum mechanics by
someone who takes off their belt or holds a filled coffee cup with one
hand and performs an elegant gyration of their arm\footnote{Some readers may have even been subjected to ``spinor spanners.''}.  It is why spinors must change sign under rotations by $2\pi$ and why the spin and statistics of particles are interrelated.  Now that we have introduced the notion of a lift when discussing the hedgehog charge, it is simplest to demonstrate this fact with yet another lift.

The Pauli matrices:
\begin{equation}
{\boldsymbol\sigma}_1=\left[\begin{matrix} 0 & 1\\ 1 & 0\end{matrix}\right],\quad
{\boldsymbol\sigma}_2 =\left[\begin{matrix} 0& -i\\ i & 0\end{matrix}\right],\quad
{\boldsymbol\sigma}_3 =\left[\begin{matrix} 1 & 0\\ 0 & -1\end{matrix}\right],
\end{equation}
satisfy ${\boldsymbol\sigma}_i{\boldsymbol\sigma}_j = i\epsilon_{ijk}{\boldsymbol\sigma}_k + \delta_{ij}\mathbf{1}$ as follows from their commutators and anti-commutators.
This can be used to form a simple way to parameterize rotations in three-dimensions.  Write any vector ${\vec x}=\left[x_1,x_2,x_3\right]$ as the matrix ${\bf x}=x_k{\boldsymbol\sigma}_k$ where we employ the summation convention over repeated indices. Then ${\bf x}^2 = \vert \vec x\vert^2 \mathbf{1}$ is the unit matrix times the squared length of the vector.  Moreover, if $\bf U$ is any $2\times 2$ unitary matrix, then if we define the similar matrix ${\bf x}'={\bf U}^\dagger {\bf x} {\bf U}$, we find that $\left({\bf x}'\right)^2 = {\bf U}^\dagger \vert\vec x\vert^2 \mathbf{1}{\bf U} = {\bf x}^2$ so under a unitary transformation the magnitude of the matrix is unchanged, precisely what we need for rotations.  We finally note that the matrix ${\bf U}({\vec\theta}) = {\bf 1} - i\frac{1}{2}\theta_k{\boldsymbol\sigma}_k$ generates infinitesimal rotations of $\bf x$ since to linear order in $\vec \theta$ (note that these angles are not the same as the Euler angles $\alpha$, $\beta$, and $\gamma$):
\begin{eqnarray}
{\bf U}^\dagger{\bf x}{\bf U} &=& \left[{\bf 1}+i\frac{1}{2}\theta_k{\boldsymbol\sigma}_k\right]x_i{\boldsymbol\sigma}_i\left[{\bf 1}-i\frac{1}{2}\theta_j{\boldsymbol\sigma}_j\right ], \nonumber\\&=&{\bf x} +\frac{i x_i\theta_k{\boldsymbol\sigma}_k{\boldsymbol\sigma}_i - ix_i\theta_k{\boldsymbol\sigma}_i{\boldsymbol\sigma}_k}{2}+ {\cal O}(\theta^2)\nonumber\\ &= & \left[x_j + x_i\theta_k\epsilon_{ikj}\right]{\boldsymbol\sigma}_j+ {\cal O}(\theta^2),
\end{eqnarray}
resulting in precisely the expression for the rotated vector.  The full rotation is\footnote{To establish this exponential formula, consider the unitary matrix $\bf S$ which diagonalizes the Hermitian matrix ${\bf t}=\theta_k{\boldsymbol\sigma}_k/2={\bf S}\mathbf{\Lambda}{\bf S}^\dagger$ where $\mathbf{\Lambda}$ is the diagonal matrix of eigenvalues.  Then ${\bf T}=\exp\{i{\bf t}\} = \exp\{i{\bf S}\mathbf{\Lambda}{\bf S}^\dagger\}={\bf S}\exp\{i\mathbf{\Lambda}\}{\bf S}^\dagger$.  Since $\hbox{Tr}\;{\mathbf{\Lambda}} = \hbox{Tr}\;{\bf t} = 0$, $\mathbf{\Lambda}=\hbox{diag}[\lambda,-\lambda]$, and it follows that 
\begin{eqnarray}
{\bf T} &=& {\bf S}\left[\begin{matrix} e^{i\lambda}&0\\ 0& e^{-i\lambda}\end{matrix}\right]{\bf S}^\dagger=
\cos(\lambda)\mathbf{1}  + i \frac{\sin(\lambda)}{\lambda}{\bf S}\mathbf{\Lambda}{\bf S}^\dagger, \nonumber\\
&=&\cos(\lambda)\mathbf{1} + i\sin(\lambda) \frac{\bf t}{\lambda}. \nonumber
\end{eqnarray}
Since $-\lambda^2=\det\mathbf{\Lambda} = \det{\bf t} = -\frac{\vert{\vec\theta}\vert^2}{4}$, the identity follows.}
\begin{equation} 
{\bf U} = e^{-i\theta_k{\boldsymbol\sigma}_k/2} = \cos\biggl(\frac{\vert\vec\theta\vert}{2}\biggr)\mathbf{1} - i\sin\biggl(\frac{\vert\vec\theta\vert}{2}\biggr)\frac{\theta_k}{\vert{\vec\theta}\vert}{\boldsymbol\sigma}_k .
\end{equation}
Thus, each rotation in $SO(3)$ can be identified with a $2\times 2$ unitary matrix with unit determinant, the group $SU(2)$.  However, both $\bf U$ and $-\bf U$ generate the {\sl same} rotation or, in other words, $\vec\theta$ and $-\vec\theta$ generate the same rotation, and thus, like the lift of $\mathbb{RP}^2$ to $\mathbb{S}^2$, we have a sign ambiguity.  Starting at $\vec\theta=0$, we can move along $\theta_1$, holding $\theta_2=\theta_3=0$.  In this case as $\theta_1$ goes from $0$ to $2\pi$, ${\bf U}(0,0,0) = -{\bf U}(2\pi,0,0)$.  We end up at the identical place in $SO(3)$, but at a different place in $SU(2)$.  It follows that this loop is not contractible to a point and so the map from the measuring circuit to $SO(3)$ which winds by $2\pi$ is one of the non-trivial elements of $\pi_1[SO(3)]$, traditionally denoted as $-{\bf 1}$.  

Dividing out by $H_b$ yields additional defects akin to the stable nematic defects of charge $1/2$.  In this case, however, there are $\pi$ rotations around each of $\hat x$, $\hat y$, and $\hat z$.  Lifting those to $SU(2)$, we find that they correspond to $\pm i{\boldsymbol \sigma}_1$,  $\pm i{\boldsymbol \sigma}_2$ and  $\pm i{\boldsymbol \sigma}_3$, respectively, where the sign ambiguity corresponds to the same lifting ambiguity as above.  These group elements have a nasty habit -- they don't commute, so a rotation of $\pi$ around $\hat x$ followed by a rotation of $\pi$ around $\hat y$, followed by a rotation of $-\pi$ around $\hat x$ gives $\left[i{\boldsymbol\sigma}_1\right]^{-1}\left[i{\boldsymbol\sigma}_2\right]\left[i{\boldsymbol\sigma}_1\right] = i{\boldsymbol\sigma}_1{\boldsymbol\sigma}_2{\boldsymbol\sigma}_1 = -i{\boldsymbol\sigma}_2$.  This means that the combination of paths on $SO(3)$ that begin and end at one of the elements of $H_b$ {\sl do not commute} and $\pi_1[SO(3)/H_b]$ is {\sl non-Abelian}.  

What about point defects in a biaxial nematic?  As it turns out, there are no point defects in the biaxial system, so $\pi_2[SO(3)/H_b]=\{1\}$ the trivial group of one element.  To see this only requires a little more work and a theorem that we will not prove but is covered in standard texts~\cite{hatcher02}, including the gold standard by Mermin~\cite{mermin79}.  First we ask what does $SU(2)$ look like as a manifold?  Fortuitously, we have the answer in the above algebra.  Note that we can consider a general matrix depending on the four real parameters $\tilde x=(x_0,\vec x)\in \mathbb{R}^4$: 
\begin{equation}
{\bf V} = x_0\mathbf{1} + ix_k{\boldsymbol\sigma}_k = \left[\begin{matrix} x_0 + i x_3 & ix_1 + x_2\cr ix_1 - x_2& x_0 - i x_3\end{matrix}\right],
\end{equation}
so $\det{\bf V} = x_0^2 + \vert\vec x\vert^2$ is the squared length of the vector $\tilde x$ in four dimensions.  Since all elements  ${\bf U}\in SU(2)$ are of this form and $\det{\bf U}=1$, we see that $SU(2)$ can be identified with the three-dimensional sphere $\mathbb{S}^3$. So what do maps from $\mathbb{S}^2$ to $\mathbb{S}^3$ look like?  They are all the same -- just as $\pi_1(\mathbb{S}^2)$ vanishes because we can shrink every loop on the sphere to a point, so too can we shrink every two-sphere on a three-sphere to a point.  Thus $\pi_2(\mathbb{S}^3) = \pi_2[SU(2)]=\{1\}$.  To calculate $\pi_2[SO[3]/H_b]$, we need only use the theorem that $\pi_2(G) = \pi_2(G/H)$ if $H$ is a discrete group\footnote{For the {\sl cognoscenti} recall that there is a long exact sequence of homotopy groups
\begin{equation*}
\ldots\rightarrow \pi_2(H) \rightarrow \pi_2(G) \rightarrow \pi_2(G/H) \rightarrow \pi_1(H) \rightarrow\ldots
\end{equation*}
and if $H$ is discrete then $\pi_2(H)=\pi_1(H) = \{1\}$, from which it follows that $\pi_2(G)=\pi_2(G/H)$.}.  Note that we saw this with $\mathbb{S}^2$ and $\mathbb{RP}^2$ -- they both shared the same $\pi_2$.  

\begin{figure}[!h]
\begin{center}
\includegraphics[width=3.25in]{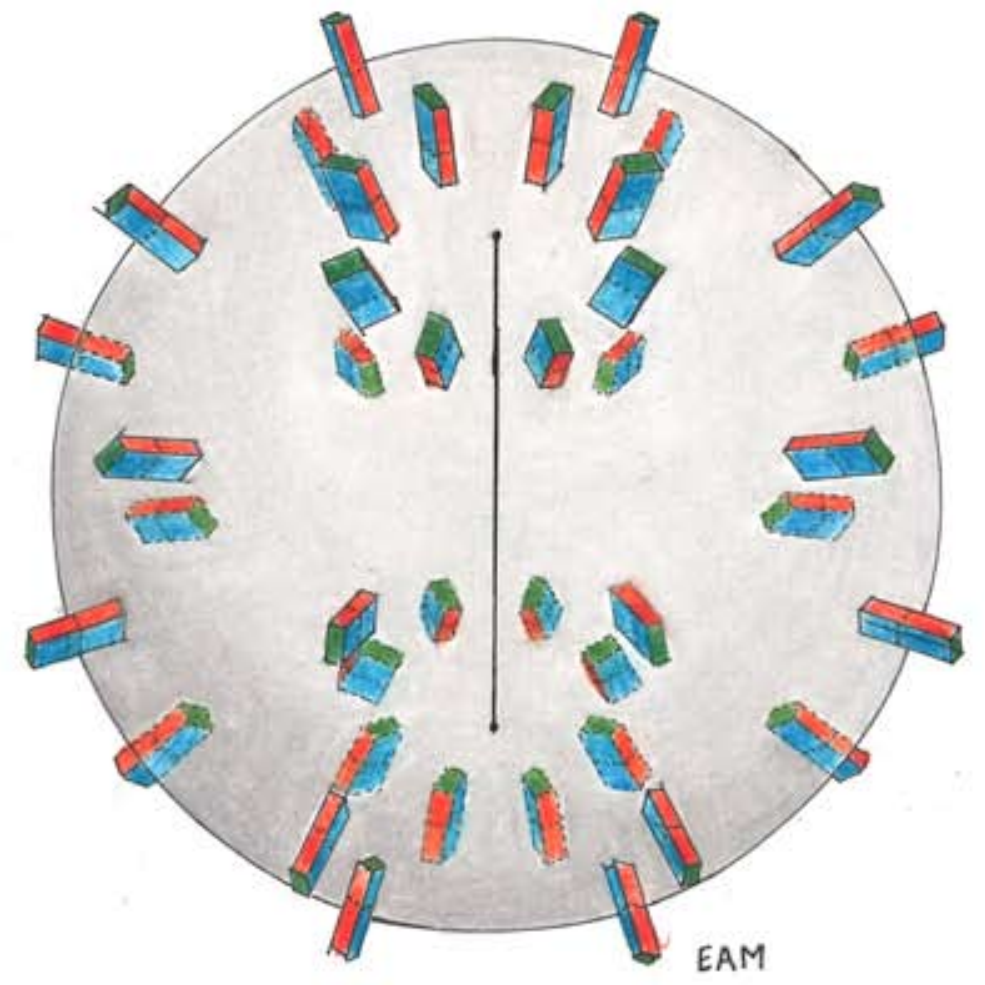}
\caption{Schematic of biaxial order on a colloid with radial anchoring conditions for the long axis. The two shorter axes undergo a $2\pi$ rotation at the North and South Poles and the particle is threaded by a $-{\bf 1}$ disclination so that a Saturn ring defect accompanying the colloid is linked with this biaxial disclination.}
\label{fig:biaxial_hedgehog}
\end{center}
\end{figure}

Precisely because the biaxial nematic does {\sl not} support point
defects, it serves to illuminate our previous discussion of
disclination loops and their relation to hedgehogs. To see what we
have gained it suffices to consider the simplest example: a
disclination loop surrounding a single colloid in a Saturn ring
configuration, illustrated in FIG. \ref{fig:biaxial_hedgehog}. As usual the meridional cycle of a torus sheathing the disclination records the non-trivial element of $\pi_1(\mathbb{RP}^2)$ and here, since the disclination is not linked with another, the longitudinal cycle records the trivial element. However, this cycle is still interesting. Note, in particular, that the director field on the inside of the torus (the part closest to the colloid) undergoes a $2\pi$ rotation as we traverse this longitudinal cycle. Thus, if we were to decorate the uniaxial texture with a perpendicular short axis to produce a biaxial nematic we would find that this cycle corresponded to a non-contractible loop in $SO(3)/H_b$, recording the element $-\bf 1$ of $\pi_1(SO(3)/H_b)$. The Saturn ring was linked with another disclination after all! An all but {\sl invisible} one only revealed by adding a small amount of biaxiality. This additional biaxial defect is not present for all defect loops: it is associated, as anticipated at the end of the last section, with a disclination loop that carries odd hedgehog charge, of which the Saturn ring is the simplest example.  

It is useful to explore this connection a little further. Suppose we remove the colloid, leaving behind just the disclination loop and the texture on the torus surrounding it. Could we fill things back in differently, say without the hedgehog? Note that the configuration on the inside of the torus is just the same as on a cylindrical capillary with perpendicular anchoring. Thus to avoid a singularity the director will have to escape in the third dimension, just as it does in a capillary. However, if we have biaxial order then the singularity cannot be avoided; although the long axis is well defined along the capillary axis the two shorter axes are not and we have a $-\bf 1$ disclination\footnote{This transferral of winding from one axis to a perpendicular one is the celebrated Mermin-Ho relation \cite{mermin76}.}. Let us assume that the texture is uniform at large distances so that this defect too closes up to form a loop. Technically, we are considering textures with defects on $\mathbb{S}^3$, the usual three-dimensional Euclidean space with a point at infinity added.  The biaxial $-\bf 1$ disclination can also be surrounded by a torus, whose meridian records a $2\pi$ winding and longitude records the $\pi$ winding of our original $1/2$ loop with which it is linked. Now by escaping in the third dimension the long axis can be made regular on any local section of the $-\bf 1$ defect, but this cannot be extended globally since the long axis is required to undergo a $\pi$ rotation along any path encircling the $1/2$ disclination. This $\pi$ rotation reverses the orientation of the director and converts an initial escape `up' into an escape `down' so that there must be a mismatch somewhere. Just as in the cylindrical capillary this mismatch marks the location of a point defect in the uniaxial nematic -- the odd hedgehog.

\section{Conclusion}
\label{sec:more}

In the hope of avoiding to provide only a poor imitation of the many excellent reviews of the homotopy theory of defects already available, {\sl e.g.} \cite{mermin79,michel80,trebin82,kurik88}, we have focused our discussion on a single physical system, nematic liquid crystal colloids, rather than provide a general survey and have eschewed, as far as possible, all of the formalities of algebraic topology. 

A recurring issue is the necessity of a base point to induce the group
structure on the set of homotopy classes. From a physical perspective the base point is really a fiction, a convenience introduced for its useful computational attributes rather than because it conveys any deep significance.  As mentioned in Section~\ref{sec:lines} the choice of base point for this purpose is entirely arbitrary so that any choice we make should yield the same results as any other. Fortunately, it is a simple exercise, proved in all of the standard reviews \cite{mermin79}, to show that the homotopy groups $\pi_k(\text{GSM},\Upsilon_0)$ and $\pi_k(\text{GSM},\Upsilon_0^{\prime})$, defined with base points $\Upsilon_0$ and $\Upsilon_0^{\prime}$, respectively, are isomorphic.  The physical properties of defects should be independent of any choice of base point that we may make and the isomorphism between homotopy groups with different base points assures this.  More importantly, however, a continuous deformation of a liquid crystal texture need not, in general, hold any point on any measuring circuit fixed while all others can freely vary.  No physical distinction should be ascribed to defects that are non-homotopic when a base point is used, but become homotopic when all points are allowed to freely vary. This is an issue for hedgehogs where, as we saw in Section~\ref{sec:hedgehogs}, dragging a point defect around a disclination line required us to carefully keep track of the base point.
Although at the end of the process the base point returned to its original value in $\mathbb{RP}^2$, the journey was not uneventful and resulted in the reversal of the sign of the hedgehog charge. Thus while $d$ and $-d$ are distinct when a base point is held fixed, they may be freely converted into each other if all points on the measuring circuit are allowed to vary.  
Without the base point the group structure is lost and since, as we have argued, it is more natural to do without the base point, {\sl the combination of defects in condensed matter does not normally follow the laws of group composition}\footnote{This is even more vivid in systems with broken translational symmetry such as smectic liquid crystals~\cite{chen09}.}; cases such as the XY or Heisenberg models are exceptions to this norm. Of course the group composition can be regained, but only at the expense of keeping track of paths that the defects move along -- the tethers of Section~\ref{sec:loops}.

Throughout this colloquium we have avoided discussing any details of how the defects themselves appear and disappear.  Indeed, when the authors have tried to observe liquid crystal textures under the microscope, we have noted, upon cooling, the sudden appearance of many defects followed by a coarsening of defects too rapid for us to nicely photograph\footnote{This may reflect more the experimental abilities of the authors than the true nature of liquid crystals.}.  The coarsening is a situation in which the defects become dynamical objects.  Not only do they move, they coalesce into defect/anti-defect pairs and disappear.  Though the addition rules afforded by the general homotopy group structure are preserved, the sample itself changes.  Since defects are places in which the local order is ill-defined, we can count the number of defects as well as measuring their charge.  When defects annihilate the defect number changes!  To account for this we have to perform ``surgery'' where we remove an arbitrarily small region of the sample along an incision and replace it with a new region that matches the texture smoothly along the cut.  A proper formulation of this problem is beyond the scope of this review and will appear elsewhere \cite{chen12}.

We have intended to bring to light the subtleties of the theory of topological defects in experimentally realizable systems. With the advent of recent experimental work on colloidal inclusions in liquid crystals, these issues are no longer purely academic.  We hope we have given the reader the tools to go forth, combine, and multiply by themselves.

\acknowledgments{We are grateful to Frederick Cohen, Simon \v{C}opar, Tom Lubensky, Carl Modes, Miha Ravnik, and Jeffrey Teo for insightful discussions.  We thank the School of Mathematics at the Institute for Advanced Study and the Aspen Center for Physics for their hospitality whilst this work was carried out. This work was supported by DMR05-47230. We dedicate this paper to the memory of Professor Yves Bouligand and to his profound contributions to the physics of ordered matter.}

\bibliographystyle{apsrmp}

\end{document}